\documentclass{jfm}

\usepackage{graphicx}
\usepackage{newtxtext}
\usepackage{newtxmath}
\usepackage{natbib}
\usepackage{caption,subcaption}
\usepackage{mathrsfs} 
\usepackage{tabularx}
\usepackage{hyperref}
\usepackage{ulem}
\usepackage{soul}
\hypersetup{
    colorlinks = true,
    urlcolor   = blue,
    citecolor  = black,
}

\newcommand{\RomanNumeralCaps}[1]

\usepackage{bm,comment}

\def\ee{\mbox{e}}
\def\dd{\mbox{d}}
\def\bea{\begin{equation}}
\def\eea{\end{equation}}

\def\hs{\omega}

\def\bu{\zh{u}}
\def\bx{\zh{x}}
\def\bul{\overline{\zh{u}}}
\def\ul{\overline{u}}
\def\fl{\overline{f}}
\def\ri{\mathrm{i}}
\def\yh{y}

\def\bea{\begin{eqnarray}}
\def\eea{\end{reqnarray}}

\def\cb{\bar{c}}

\def\bea{\begin{eqnarray}}
\def\eea{\end{eqnarray}}
\newcommand\Ca{\mbox{\textit{Bo}}}  

\newcommand{\zh}{\boldsymbol}
\newcommand{\mylab}[1]{\label{#1}}

\definecolor{cadmiumgreen}{rgb}{0.0, 0.42, 0.24}

\def\cbl{\color{black}}
\def\cb{\color{black}}

\title[On the transition to dripping of an inverted liquid film]{On the transition to dripping of an inverted liquid film}

\author[M. G. Blyth, T.-S. Lin and D. Tseluiko]%
{M. G. Blyth$^1$, \ns
T.-S. Lin$^2$, \ns
D. Tseluiko$^3$
}

\affiliation{
$^1$School of Mathematics, University of East Anglia, Norwich, NR4 7TJ, UK\\[\affilskip]
$^2$Department of Applied Mathematics, National Yang Ming Chiao Tung University, Hsinchu 30010, Taiwan\\[\affilskip]
$^3$Department of Mathematical Sciences, Loughborough University,
Loughborough,~LE11~3TU,~UK\\[\affilskip]
}

\pubyear{2019}
\volume{000}
\pagerange{000--000}
\date{?; revised ?; accepted ?. - To be entered by editorial office}

\begin{document}

\maketitle
\begin{abstract} 
The transition to dripping in the gravity-driven flow of a liquid film under an inclined plate is investigated at zero Reynolds number. Computations are carried out on a periodic domain assuming either a fixed fluid volume or a fixed flow rate for a hierarchy of models: two lubrication models with either linearised curvature or full curvature (the LCM and FCM, respectively), and the full equations of Stokes flow. Of particular interest is the breakdown of travelling-wave solutions as the plate inclination angle is increased. For any fixed volume the LCM reaches the horizontal state where it attains a cosine-shaped profile. For sufficiently small volume, the FCM and Stokes solutions attain a weak Young-Laplace equilibrium profile, the approach to which is described by an asymptotic analysis generalising that of \cite{kalliadasis1994drop} for the LCM. For large volumes, the bifurcation curves for the FCM and Stokes model have a turning point so that the fully inverted state is never reached. For fixed flow rate the LCM blows up at a critical angle that is well predicted by asymptotic analysis. The bifurcation curve for the FCM either has a turning point or else reaches a point at which the surface profile has an infinite slope singularity, indicating the onset of multi-valuedness. The latter is confirmed by the Stokes model which can be continued to obtain overturning surface profiles. Overall the thin-film models either provide an accurate prediction for dripping onset or else supply an upper bound on the critical inclination angle.
\end{abstract}
\vspace{-0.4cm}

\section{Introduction}\label{sec:intro}

The flow of a  viscous liquid film down an inclined plate is of fundamental theoretical and experimental interest (see, for example, the review article by \citet{craster2009dynamics} and the monograph by \citet[][]{kalliadasis2011falling}.
While most research on this topic has focused on the flow down the upper side of a plate, there has been steadily growing interest in understanding the dynamics of inverted films, including those flowing down the underside of a plate, and hanging beneath an inverted plane wall. Of particular interest are the fundamental mechanisms responsible for dripping wherein gravitational effects lead to the formation of large amplitude structures on the surface of the film which form into droplets and ultimately detach \citep{indeikina,Lin_Kondic_2010,brun2015,Scheid_etal_Phys_Fluids_28_2016,Rohlfs_etal_2017,Kofman_etal_2018,Charogiannis,zhou2022dripping}.

A wide range of technological devices and industrial, environmental and biomedical processes make use of inverted or partially inverted films. Industrial applications include liquid film coating and  fills in cooling towers \citep{Rohlfs_etal_2017}, and environmental applications include glacier hydrology and the morphogenesis of cave patterns \citep{Camporeale_2017}. In biomedicine inverted films arise in the process of microbicide  epithelial coating used for protection against HIV \citep{Hu_2016}. Understanding the dynamics of inverted films, including dripping phenomena, is also relevant to forensic science, for example in blood pattern analysis \citep{kabaliuk2013blood}. 
In these and other applications, various fluid dynamical features are of especial importance. For example, destabilising effects such as gravity and inertia produce spatial heterogeneity that might be desirable in some applications (e.g.\ heat transfer in cooling films and falling-film chemical reactors) but detrimental in others (e.g.\ in film coating). 

A viscous film falling down a vertical wall under the influence of gravity is stable in the 
absence of inertia \citep{benjamin1957wave,yih}. As the inclination angle is increased beyond the vertical, the film becomes partially inverted and thereby susceptible to the gravitational Rayleigh-Taylor linear instability. If the inclination angle is not too great one might expect that 
linear disturbances will grow, become dominated by nonlinear effects, and ultimately saturate; and indeed experiments \citep{brun2015} show that the film surface develops large amplitude travelling-wave structures. 
However, common experience suggests that beyond a critical angle  
disturbances do not saturate but continue to grow in size, eventually leading to 
dripping. This suggests that a possible approach to understanding and predicting dripping onset is to track the branch of travelling-wave solutions via a continuation method with a view to observing some form of breakdown at the critical angle. A preliminary attempt at this for the fixed volume case was made by \cite{Kofman_etal_2018} using weighted residual integral boundary layer models.

In the fully inverted state there is no preferential flow direction and it is possible to obtain static 
equilibria that correspond to solutions of the Young-Laplace (henceforth YL) equation expressing a balance 
between surface tension and gravity. Such equilibria were constructed in two dimensions by 
\cite{pitts1973stability}, who showed that in agreement with physical intuition static solutions 
exist provided that the drop volume is sufficiently small. In line with this the branch of static 
solutions computed by \cite{pitts1973stability} has a turning point at a certain volume; and, in 
fact, two possible static equilibria co-exist over a particular range of volumes, although only one 
of these is stable \citep{pitts1973stability,lowry1995capillary}. Static hanging-drop solutions 
have also been computed for inclined planes \citep{pozrikidis2012stability}. YL 
solutions are relevant to the dripping problem since, following the travelling-wave protocol 
suggested above and provided that certain conditions are met, they should be attained in the 
limit as the plate becomes fully inverted. This will happen in the fixed volume case if the fluid volume is less than the threshold value identified by \cite{pitts1973stability}.
However, if the volume exceeds this threshold value or, as is intuitively clear, if the flow rate in the film is fixed, then the horizontal state cannot be reached via continuation. In this case
we might anticipate that the branch of travelling-wave solutions cannot be continued to the fully inverted state and, consequently, that dripping should occur at some angle before this.

In an alternative viewpoint, the onset of dripping on an inverted film has been discussed in the context of convective and absolute instability by a number of authors including \cite{brun2015}, who proposed the idea, \cite{Scheid_etal_Phys_Fluids_28_2016},  \cite{Kofman_etal_2018}, and also \cite{tomlin2020instability} who applied it to an electrified film.
This approach was criticised by \cite{zhou2022dripping}, who claimed that the dripping 
mechanism is intrinsically nonlinear and, consequently, cannot be adequately explained as a 
transition from convective to absolute instability.
In an approach similar to that adopted here but for the fixed volume case only, \cite{zhou2022dripping} carried out numerical computations of the full Navier-Stokes equations using the open-source software Basilisk (http://basilisk.fr). By solving the unsteady form of the equations from a prescribed initial condition, and by slowly and continuously increasing the inclination angle of the plate during the simulation, they were able to compute near travelling-wave states in which the film exhibits a localised drop-like bulge in the centre of the computational domain. In particular, they found that such a state is reached provided that the inclination angle does not become so large that the localised drop detaches from the film, corresponding to dripping. They linked the drop detachment process with the point at which the curvature at the drop tip exceeds the tip curvature of a static YL drop of maximum volume for a fully inverted plate. They provided corroborating evidence to support this connection by overlaying the YL solutions onto their simulated wave profiles at times near to the onset of dripping.

\cb There remain two outstanding issues of primary importance. The first is to provide an indicator for the onset of dripping that can be easily measured or computed and hence utilised in practice by the wider community. The second
is a rigorous mathematical justification of the use of this indicator. \cb In an attempt to provide these, 
in this paper we compute  travelling-wave branches both for the fixed volume case and for the fixed flow rate case. 
Both of these set-ups are relevant to applications.
In particular we perform a proper continuation study of travelling-wave solutions supported by asymptotic analysis. For simplicity, and to 
focus on the competition between surface tension and gravity, we disregard fluid inertia. 
We study models with increasing levels of complexity. At the simplest level we employ a classical lubrication model 
that includes a rational linearisation of the curvature at the film surface. This is complemented 
by an ad hoc generalised model in which the same equation is used but with the full curvature 
term substituted arbitrarily. 
\cbl
Such an approach has been followed by other authors in the 
literature for various related problems 
\cite[e.g.][]{eggers2008physics, lopes2018multiple, Kofman_etal_2018}. 
Here this step is motivated by the observation that for inverted flow we expect large-amplitude surface 
deformations and, consequently, that the capillary pressure, and hence the curvature, will play a dominant 
role in the dynamics. Moreover, including the full curvature term allows for the full Young-Laplace equation 
for the static configuration to be recovered in the limit when the plate tends to become horizontal.
\cb

The rest of the paper is organised as follows. In \S2 we define the mathematical problem and 
discuss the relevant dimensionless parameters. In \S3 we present the thin-film equations and 
discuss an equivalence between the fixed flow-rate and fixed volume cases that holds for the 
linearised curvature model. 
In \S4 we present 
an asymptotic analysis of the full curvature lubrication model for the fixed volume case that is valid in the limit as the plate becomes 
fully inverted. The numerical method that we use for the full Stokes computations, which is based on a boundary-integral formulation and which employs a spectrally-accurate representation for the flow, is developed in \S5. In \S6 we present the results of our numerical computations. Finally, in \S7 we summarise and discuss our findings.

\section{Problem statement}\label{sec:problem}

We consider an inverted two-dimensional viscous liquid film that is flowing down the underside of a plane wall that is inclined at
an angle $\beta$ to the horizontal, where $\pi/2\leq \beta\leq \pi$ (see figure~\ref{fig1}a). 
We use Cartesian coordinates $(\tilde x,\tilde y)$ with $\tilde x$
and $\tilde y$ measuring distance along the wall and normal to it, respectively, as shown in the figure. \cbl We use tildes to indicate dimensional variables. \cb Assuming that
 inertia is negligible, the flow is governed by the Stokes momentum and continuity equations, namely
\begin{eqnarray}\label{NS}
\zh{0} = - \widetilde \nabla \tilde p  + \mu \widetilde \nabla^2 \tilde \bu + \rho \,\zh{\widetilde G}, \qquad \widetilde \nabla \cdot \tilde  \bu = 0,
\end{eqnarray}
where $\tilde p$ and $\tilde \bu=(\tilde u,\tilde v)$ are respectively the pressure and velocity in the liquid film, $\zh{\widetilde G} = g (\sin\beta, -\cos \beta)$ is the gravitational acceleration, and $\widetilde \nabla = (\partial/\partial \tilde x,\partial/\partial \tilde y)$. The dynamic viscosity and density of the fluid are $\mu$ and $\rho$, respectively. The pressure in the air below the film is taken to be zero without loss of generality. 

At the wall we have the no-slip condition, $\tilde \bu = \zh{0}$ at $\tilde y=0$. At the free surface we must impose the kinematic condition,
$D\tilde f/D\tilde t = 0$,
where $\tilde f=0$ describes the location of the free surface and $\tilde{t}$ denotes time. In the simplest case the free surface is a graph of a function such that $\tilde f=\tilde y-\tilde h(\tilde x,\tilde t)$ and the kinematic condition takes the form
\begin{eqnarray}
\label{banana2}
\tilde v=\tilde h_{\tilde t} + \tilde u\tilde h_{\tilde x},
\end{eqnarray}
at $\tilde y=\tilde h(\tilde x,\tilde t)$. Also at the free surface we impose the dynamic stress conditions
\begin{eqnarray}
\label{grapes}
\zh{t}\cdot \zh{\widetilde T} \cdot \zh{n} = 0,
\qquad
\zh{n}\cdot \zh{\widetilde T} \cdot \zh{n} = \sigma \tilde \kappa,
\end{eqnarray}
where $\sigma$ is the surface tension coefficient, and $\zh{t}$ and $\zh{n}$ are the unit tangent and unit normal vectors at the free surface, respectively, with $\zh{n}$ pointing into the liquid. The free surface curvature is given by $\tilde \kappa=\widetilde \nabla\cdot \zh{n}$ and is positive/negative as illustrated in 
figure~\ref{fig1}(a), and {\color{black} $\zh{\widetilde T}=-\tilde p\zh{I} + \mu (\widetilde \nabla \tilde\bu + \widetilde \nabla \tilde\bu^T)$} is the Newtonian stress tensor in the liquid (the superscript $T$ denotes matrix transpose).

For a flat film of uniform thickness $h_0$ the velocity field inside the film adopts the unidirectional Nusselt velocity profile with velocity components
\bea \label{nusselt}
\tilde u(\tilde y) = \frac{\rho g \sin \beta}{2\mu}(2h_0\tilde y-\tilde y^2), \qquad \tilde v = 0.
\eea
The associated flow rate is
\bea \label{flowrate}
\tilde q = \frac{\rho g h_0^3 \sin \beta}{3\mu}.
\eea
%
\begin{figure}
\centering
{\includegraphics[width=5in]{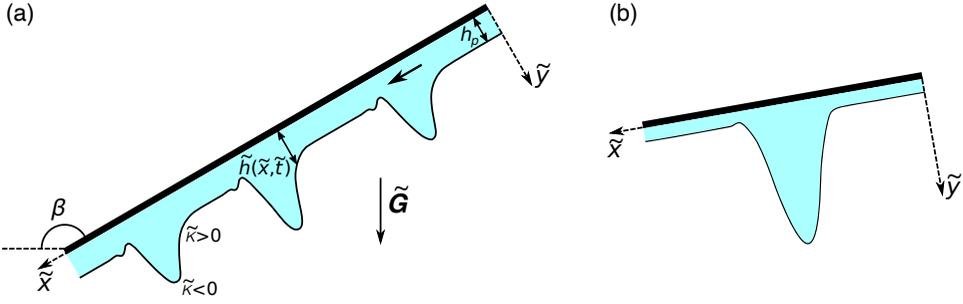}}
\caption{(Color online) (a) Schematic representation of a viscous liquid film flow down an inclined plane wall. (b) Schematic for the fixed volume case showing a drop on a thin precursor film.
}
\mylab{fig1}
\end{figure}
When $\beta=\pi$ the film is static and the relevant solutions correspond to a balance between surface tension
and gravity, that is solutions to the YL equation discussed in Appendix~\ref{sec:static}.

According to the definition of $\tilde q$, to maintain the same flow rate for a flat film, the film thickness changes with the inclination angle so that 
\bea
h_0(\beta) = \frac{h^*}{(\sin\beta)^{1/3}}, 
\eea
where $h^*=(3\mu \tilde q/\rho g)^{1/3}$ is the flat film thickness that is obtained on a vertical wall so that $h^*=h_0(\pi/2)$. It is convenient to introduce dimensionless variables that are independent of $\beta$. To do this we use $h^*$ as the length scale, $2\mu/\rho g h^*$ as the time scale, and $\rho g h^*/2$ as the pressure scale. This reveals the dynamical importance of the dimensionless Bond number,
\begin{equation}\label{cawe}
\Ca = \frac{\rho g {h^*}^2}{2\sigma},
\end{equation}
and the dimensionless flow rate 
\bea \label{nondimq}
q = \left(\frac{2\mu}{\rho g {h^*}^3}\right) \tilde q = \frac{2}{3} h_p^3\sin \beta,
\eea
where $\tilde q$ is the dimensional flow rate for a Nusselt film defined in \eqref{flowrate}, and
\bea\label{nondimh}
h_p = \frac{h_0}{h^*}=\frac{1}{(\sin\beta)^{1/3}}
\eea
is the dimensionless upstream film thickness. {\cbl Henceforth we drop the tildes to indicate dimensionless variables.}  In what follows we shall perform calculations assuming a dimensionless fixed flow rate, $q$, at different inclination angles $\beta$; and also calculations in which the flow is assumed to be periodic in $x$ with a fixed volume in each period, again for different inclination angles. 

Simplifications can be made in the case when the thickness is small in comparison to the typical length scale of any streamwise variations. This is considered in the next section.

\section{Thin film analysis}\label{sec:long}

Assuming that the lengthscale of the interfacial deformation 
$\lambda$ is large compared with $h^*$ (i.e.\ the so called thin-film parameter $\epsilon=h^*/\lambda$ is small), the film thickness satisfies the model equation, 
made dimensionless according to the scales given in \S\ref{sec:problem},
\begin{equation}
h_t + q_x = 0, \qquad q = h^3 P_x, \qquad
P = \frac{2\sin\beta}{3}x - \frac{2\cos\beta}{3} h +\frac{1}{3\Ca} h_{xx},
\mylab{eq:TFE1}
\end{equation}
\cbl where $h(x,t)$ and $q(x,t)$ are the dimensionless film thickness and flow rate, respectively, and $P(x,t)$ represents a combination of the leading-order effects of gravity and surface tension. \cb The latter includes the hydrostatic pressure due to gravity, represented by the first two terms (giving the $x$ and $y$ components, respectively), and the capillary pressure due to surface tension, represented by the third term. Equation \eqref{eq:TFE1} is derived using systematic asymptotics in \cite{kalliadasis2011falling}  \cite[see
also][]{tseluiko2013stability}, and is valid in the present case of negligible inertia provided
that $\Ca=O(\epsilon^2)$ and $\cot \beta=O(\epsilon^{-1})$ \cite[][]{blyth2018two}. 

We seek travelling-wave solutions in the form of localised pulses or drops such that the film thickness $h$ becomes constant 
upstream and downstream. Introducing a moving-frame coordinate via the 
mapping $x\mapsto x+ c t$, for constant wavespeed $c$, a travelling wave solution $h(x)$ to (\ref{eq:TFE1}) 
must satisfy
\begin{equation}
\left[ -c h + h^{3} \left( \frac{2\sin \beta}{3} - \frac{2\cos\beta}{3}  h' +\frac{1}{3\Ca} h''' \right)
\right]'=0,
\mylab{eq:TFE2}
\end{equation}
where a prime indicates differentiation with respect to $x$.
We either fix the dimensionless flow rate, $q$, which effectively sets the film thickness upstream and downstream via 
\eqref{nondimh}, or else we fix the volume of fluid over a specified domain $[-L,L]$, that is we set 
\begin{eqnarray}\label{mass}
\int_{-L}^L h \,\mbox{d} x = V, 
\end{eqnarray}
for some $V$.

There exists an equivalence between a fixed-$V$ volume localised droplet solution with a precursor film thickness $h_p(\beta)$ for an angle $\beta$ and a fixed-$q$ flow rate solution with upstream film thickness $\hat h_p(\hat \beta)$ for a certain angle $\hat \beta$. This equivalence is established by noting that transforming \eqref{eq:TFE2} so that
\bea \label{scalings}
h \mapsto h_p(\beta) h, \quad x \mapsto (-2 \Ca\cos \beta)^{-1/2}x, \quad c \mapsto \tfrac{1}{3} (-8\Ca \cos^3\beta)^{1/2}h_p^3(\beta)c,
\eea
for the fixed volume case, and
\bea
h \mapsto \hat h_p(\hat \beta) h, \quad x \mapsto (-2 \Ca\cos \hat \beta)^{-1/2}x, \quad c \mapsto \tfrac{1}{3} (-8\Ca \cos^3\hat \beta)^{1/2}\hat h_p^3(\hat \beta)c
\eea
for the fixed flow rate case leads to the same equation, namely
\begin{equation}
\left[-c h + h^{3} \left( \gamma + h' + h''' \right)
\right]'=0,
\mylab{eq:TFE3}
\end{equation}
subject to the condition that $h$ approaches unity in the far-field, where
\bea \label{gamparam}
\gamma = - \frac{\tan \beta}{h_p(\beta)(-2\Ca \cos \beta)^{1/2}} = - \frac{\tan \hat \beta}{\hat h_p(\hat \beta)(-2Bo \cos \hat \beta)^{1/2}}>0.
\eea

The upstream film thickness for the fixed flow rate solution is known via \eqref{nondimq} to be $\hat h_p(\hat \beta)=(3q/2\sin \hat\beta)^{1/3}$, but the precursor film thickness $h_p(\beta)$ for the fixed volume localised droplet case must be found as part of the solution to the problem. Equation (\ref{eq:TFE3}) was also derived and analysed by \cite{kalliadasis1994drop} in a different context, namely drop formation on vertical fibres. They showed that solutions to \eqref{eq:TFE3} with $h(\pm \infty)=1$ blow up as $c\to \infty$ such that $\gamma \rightarrow \gamma_c\approx 0.5960$. The problem was later revisited by
\cite{yu2013velocity}, who supplied higher order corrections, noting that $\gamma = \gamma_c + 0.33c^{-2/3} + 0.19c^{-1} + O(c^{-4/3})$. The implications for the present work are: (i) for the fixed-$V$ volume case,
\bea\label{hpexp}
h_p(\beta) = \frac{\pi-\beta}{\gamma_c(2\Ca)^{1/2}} + o(\pi-\beta)
\eea
as $\beta\to \pi-$, and (ii) for the fixed-$q$ flow rate case there is a critical angle, $\hat \beta_c$ say, at which blow-up occurs, that satisfies
\bea\label{hpexp2}
 \frac{(\sin \hat\beta_c)^{4/3}}{2^{1/6}3^{1/3}q^{1/3}\Ca^{1/2}(-\!\cos \hat \beta_c)^{3/2}} = \gamma_c \approx 0.5960. 
\eea

A common approximation adopted in the literature is to replace the simplified curvature term in \eqref{eq:TFE1}
with its exact form. Such an approximation is {\it ad hoc} but it has nevertheless been used successfully to predict
thin film flows in a number of different contexts, for example 
in liquid-film breakup \cite[][]{gauglitz1988extended}. In this case the thin film system takes the form \eqref{eq:TFE1} but with 
\bea
P = \frac{2\sin\beta}{3}x - \frac{2\cos\beta}{3} h +\frac{1}{3\Ca} \kappa,
\label{eq:TFE2_full} 
\eea
where
\bea \label{kappadef}
\kappa = \frac{h_{xx}}{(1+h_x^2)^{3/2}}
\eea
is the curvature.
We will refer to this as the full curvature model (FCM), and we will refer to equation \eqref{eq:TFE1} as the linearised curvature model (LCM). Note that there is no equivalence between the cases of fixed flow-rate and fixed volume for the FCM. Similarly there is no such equivalence for the full Stokes system described in \S\ref{sec:problem}.

In the fixed-volume case our particular interest is in the limit when $\beta\to \pi-$ so that the wall tends to become horizontal. 
Numerical computations, that will be discussed in detail later, indicate that solutions take the form of localised drops with a very thin precursor film on either side, as is sketched in figure~\ref{fig1}(b). As $\beta$ approaches $\pi$, the precursor film thickness tends to zero and the drop profiles converge to solutions of the YL equation
for a static drop that represent a balance between surface tension and gravity. Such solutions are discussed in Appendix~\ref{sec:static}.

{\color{black} Given the aforementioned restrictions, the thin-film models formally break down when $\beta$ is sufficiently 
close to $\pi$. We will subsequently carry out computations for the full equations of Stokes flow with a view to assessing the performance of the thin film models. This comparison will be presented along with all of our main results in \S\ref{sec:numerics}. \cb

\section{Asymptotics for $\beta \to \pi-$ for fixed volume for the FCM}\label{appendix:asymp}

In this section we present an asymptotic analysis of the fixed volume FCM solutions in the limit $\beta\to \pi-$. According to the discussion in Appendix~\ref{sec:static}, such an analysis is relevant provided that $\Ca V \leq 2.60$ and there exists a limiting static solution (see in particular figure \ref{fig:YLsoln}). We do not attempt a similar analysis for fixed flow rate since the numerical continuation studies to be presented  in a later section indicate either the presence of a turning point or an infinite-slope singularity meaning that the angle $\beta=\pi$ is not reached. For the LCM model, the fixed volume and fixed flow rate cases are equivalent, as was noted in \S\ref{sec:long}, and the analogous asymptotic analysis has been carried out elsewhere \cite[see][]{yu2013velocity,kalliadasis1994drop}.

In a frame of reference travelling at speed $c$ in the positive $x$ direction, writing $z=x-ct$, the FCM \eqref{eq:TFE2_full}  becomes
\bea
-ch_z + q_z = 0, \qquad q = h^3 \left(\frac{2\sin\beta}{3} - \frac{2\cos\beta}{3} h_z +\frac{1}{3\Ca} \kappa_z\right),
\label{eq:app1} 
\eea
where the curvature is
\bea \label{app:kappadef}
\kappa = \frac{h_{zz}}{(1+h_z^2)^{3/2}},
\eea
assuming that the solution is a single-valued function of $z$. Integrating once we have
\bea \label{app:Qeq}
-ch + q = Q
\eea
for constant $Q$.
We introduce the parameter $\delta = \pi-\beta$ and henceforth assume that $\delta\ll 1$. The asymptotic solution has the structure depicted in figure \ref{fig:asymp} and incorporates four regions: the main part of the drop ($R_2$), the left-side matching zone (region $B_1$), the right-side matching zone (region $B_2$) and the precursor films (regions $R_1$ and $R_3$). 
%
\begin{figure}
\centering
{\includegraphics[width=3.5in]{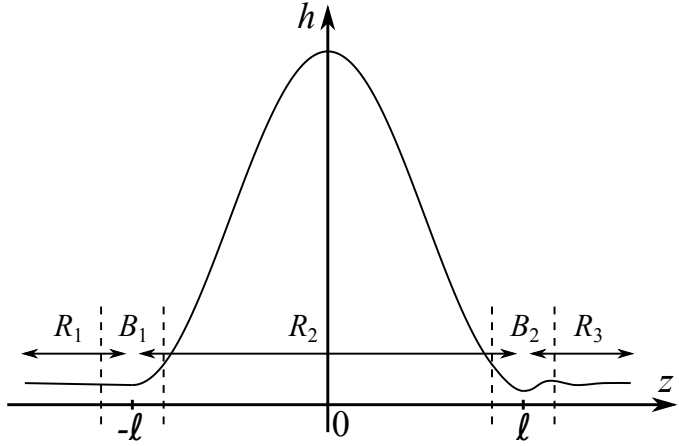}}
\caption{Sketch of the asymptotic regions for the case of fixed volume.}
\mylab{fig:asymp}
\end{figure}

In region $R_2$ we expand by writing 
\bea \label{app:hexpand}
h=h_0(z) + \delta h_1(z) + \delta^{3/2}h_2(z) + \cdots, \quad c = \delta^{3/2}c_0 + \cdots,
\quad Q = \delta^{5/2} Q_0 + \cdots,\,\,
\eea
where the forms of the expansions have been selected to allow a consistent match between the regions. The inherent degeneracy in the problem due to a translational invariance in $z$ is removed by pinning the drop with its maximum at the origin so that $h'(0)=0$, where a prime denotes differentiation with respect to $z$. 

Substituting \eqref{app:hexpand} into \eqref{app:Qeq} and integrating the leading-order equation once, we obtain
\begin{equation}
2 h_0 + \frac{1}{\Ca}\frac{h_{0}''}{(1+h_0'^2)^{3/2}} = P_0.
\end{equation}
The solution $h_0(z)$ is a static-drop at $\beta=\pi$ with the support from $-\ell$ to $\ell$. It touches the wall with zero slope at the ends, so that $h_0(\pm \ell) = h_0'(\pm \ell)=0$, and has volume $V$ so that
\bea
\int_{-\ell}^\ell h_0(z)\,dz = V,
\eea
and the volume contained in each of $h_1$, $h_2$, etc. is zero.
This solution is analysed in Appendix~\ref{sec:static}.
With the pinning condition $h_0'(0)=0$ we have
\bea
h_0 \sim a_0 (z \pm \ell)^2 \quad \mbox{as $z\to \mp \ell^{\pm}$}, \qquad
\eea
where the coefficient $a_0$ can be estimated from the numerical solution and depends on~$V$. The solution for the case $V=8$ and $\Ca=0.3$ is shown in the left panel of figure~\ref{fig:h1op} and is such that $\ell=3.3598$ and $a_0=0.3572$.

At $O(\delta)$ we find after one integration,
\bea \label{app:h1eq}
(2\Ca) h_{1}' + \left( \frac{h_1'}{(1+h_0'^2)^{3/2}} \right ) ''  = - 2\Ca.
\eea
The solution can be written in the form
\bea
h_1(z) = b_1 h_{1E}(z) + b_2h_{1O}(z) + b_3 + h_{1P}(z),
\eea
for constants $b_1$, $b_2$, $b_3$, where $h_{1E}(z)$ and $h_{1O}(z)$ are even and odd functions, respectively, and the particular integral $h_{1P}(z)$ is odd. 
We assume without loss of generality $h_{1E}(0)=1$, $h_{1O}'(0)=1$ and $h_{1P}'(0)=1$. 
Numerically computed solutions for $h_{1O}(z)$ and $h_{1P}(z)$ are shown in figure~\ref{fig:h1op}. Restricting attention to the range $[0,\ell]$, since $h_0(z)$ is symmetric about the inflection point at $z=\ell/2$ (see Appendix~\ref{sec:static}), $h_{1O}(z)$ is also symmetric about the line $z=\ell/2$. 
The pinning condition $h_1'(0)=0$ demands that $b_2=-1$.
Hence
\bea \label{app:h1Leq}
h_1(\ell) - h_1(-\ell) = 2h_{1P}(\ell).
\eea
From our numerical solution we determine that $h_{1P}(\ell) = -3.3740$.
\begin{figure}
\begin{subfigure}{0.5\textwidth}
\includegraphics[scale=0.385]{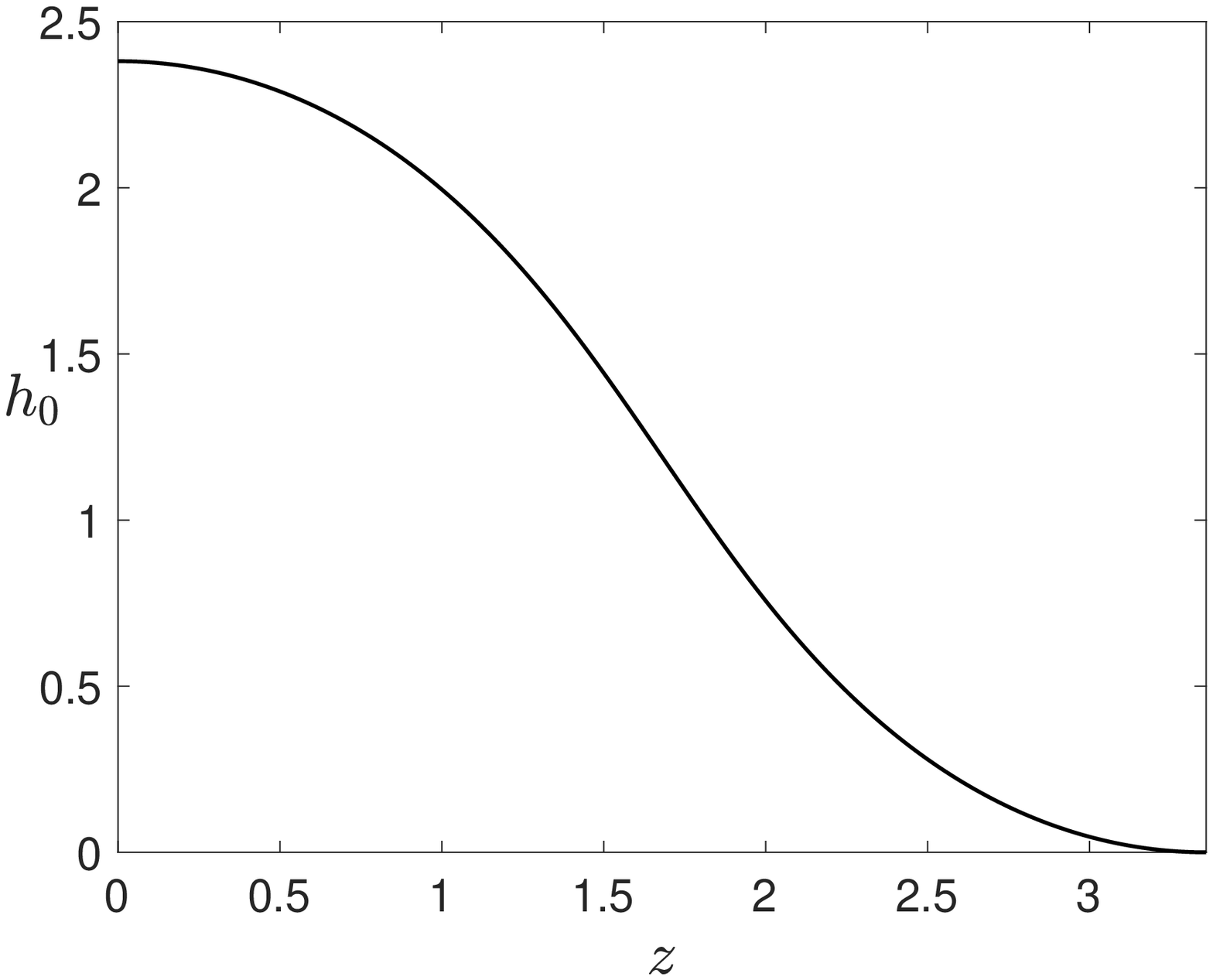}
\caption{}
\end{subfigure}
\hspace{0.1in}
\begin{subfigure}{0.5\textwidth}
\includegraphics[scale=0.385]{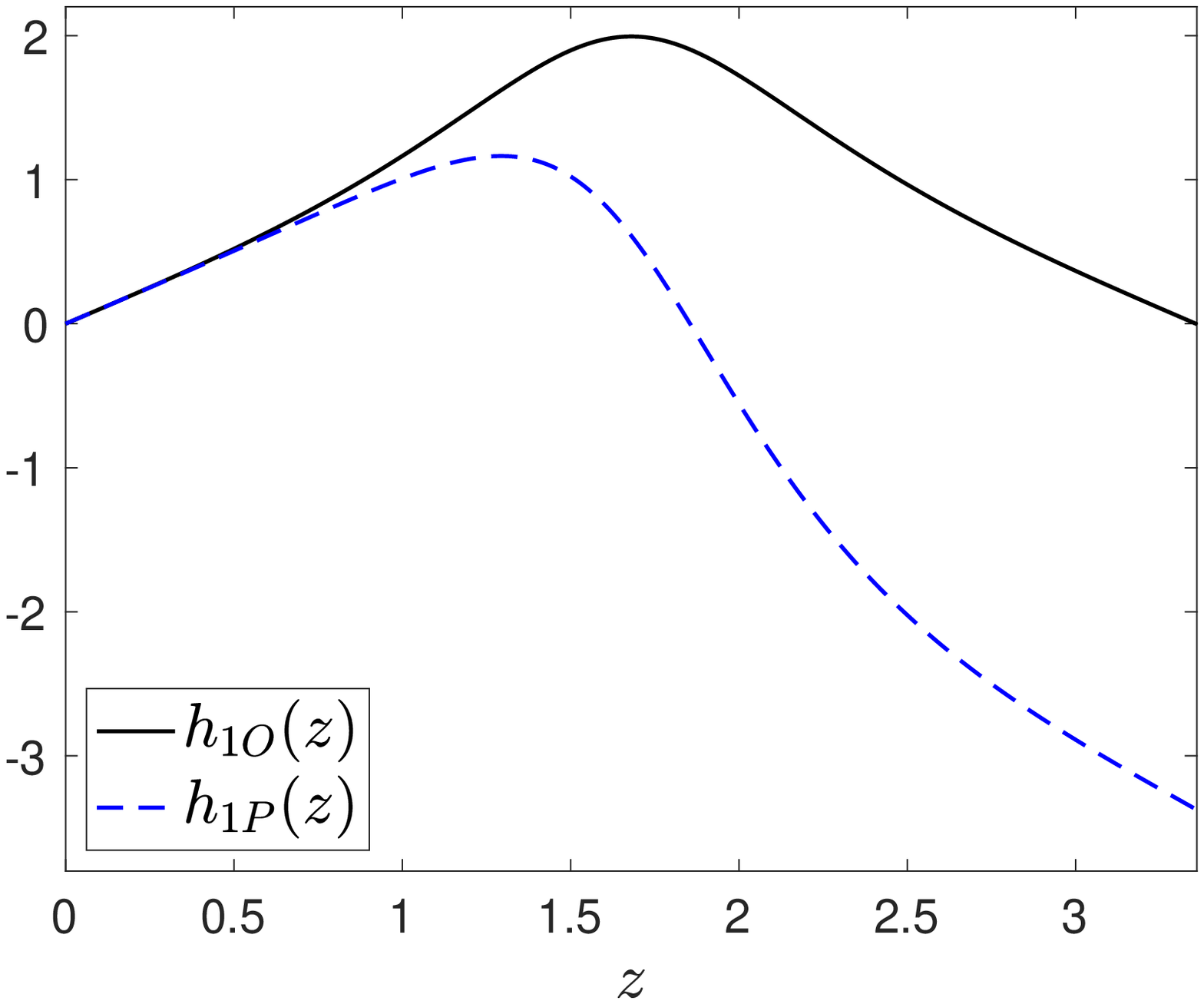}
\caption{}
\end{subfigure}
\caption{(Color online) $V=8$ and $\Ca=0.3$: (a) Leading order solution $h_0(z)$. (b) The functions $h_{1O}(z)$ and $h_{1P}(z)$.}
\mylab{fig:h1op}
\end{figure}

At $O(\delta^{3/2})$ we find
\bea
(2\Ca) h_2' + \left(  \frac{h_2'}{(1+h_0'^2)^{3/2}} \right )'' = \frac{3c_0 \Ca}{h_0^2}.
\eea
The solution has the singular behaviour 
\bea
h_2 \sim -\frac{c_0\Ca}{2a_0^2}(z\pm \ell)^{-1}
\eea
as $z\to \mp \ell^{\pm}$ signalling a breakdown in the expansion \eqref{app:hexpand} in region $R_2$ where $z\pm \ell = O(\delta^{1/2})$ thus necessitating the regions $B_1$ and $B_2$ to be discussed below.

In regions $R_1$ and $R_3$ we expand by writing
$$
h = \delta h_{p0} + \cdots.
$$
Substituting into \eqref{app:Qeq} we obtain at leading order $-c_0 h_{p0} = Q_0$.

%
\begin{figure}
\begin{subfigure}{0.5\textwidth}
\includegraphics[width=2.6in]{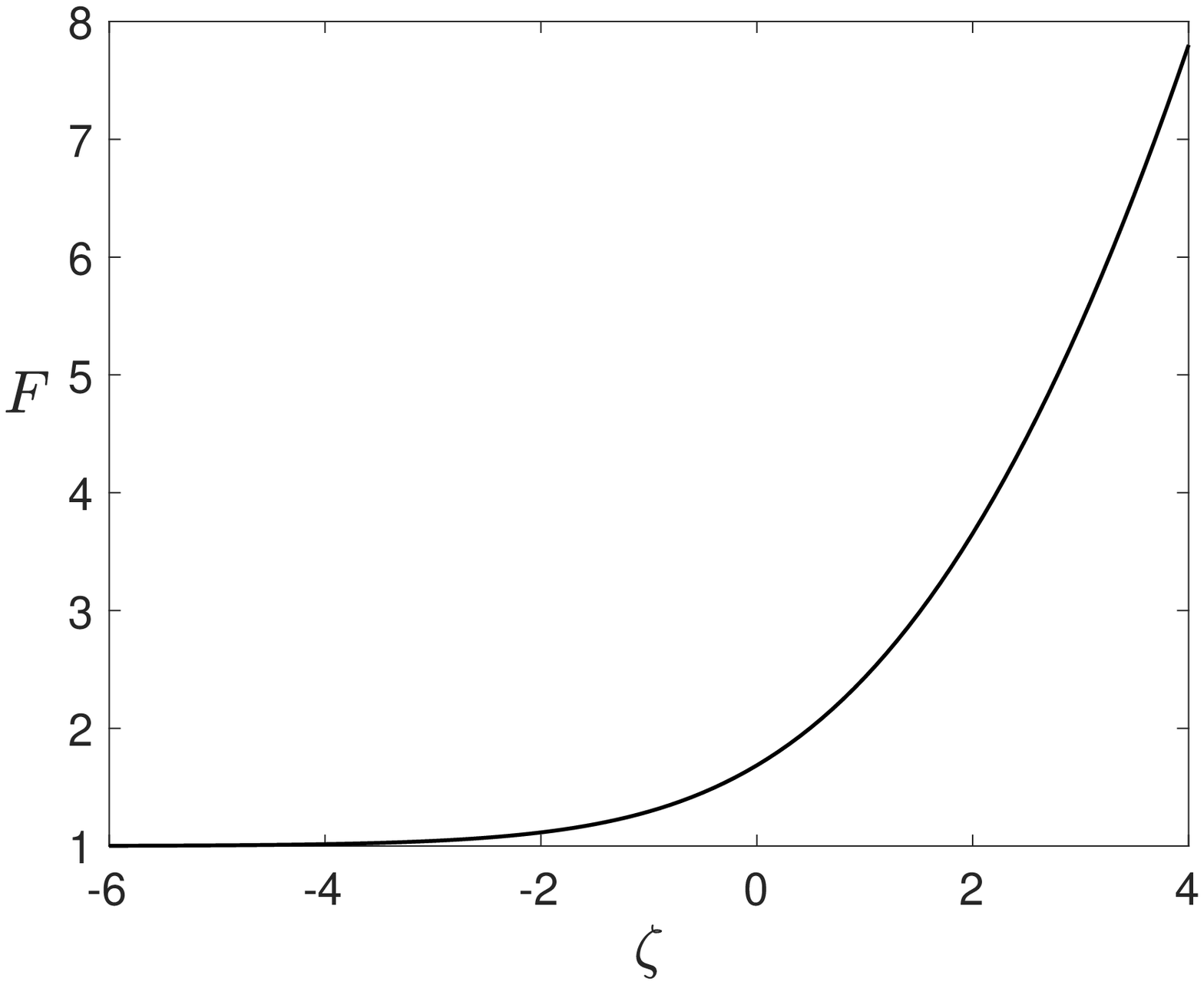}
\caption{}
\end{subfigure}
\begin{subfigure}{0.5\textwidth}
\includegraphics[width=2.6in]{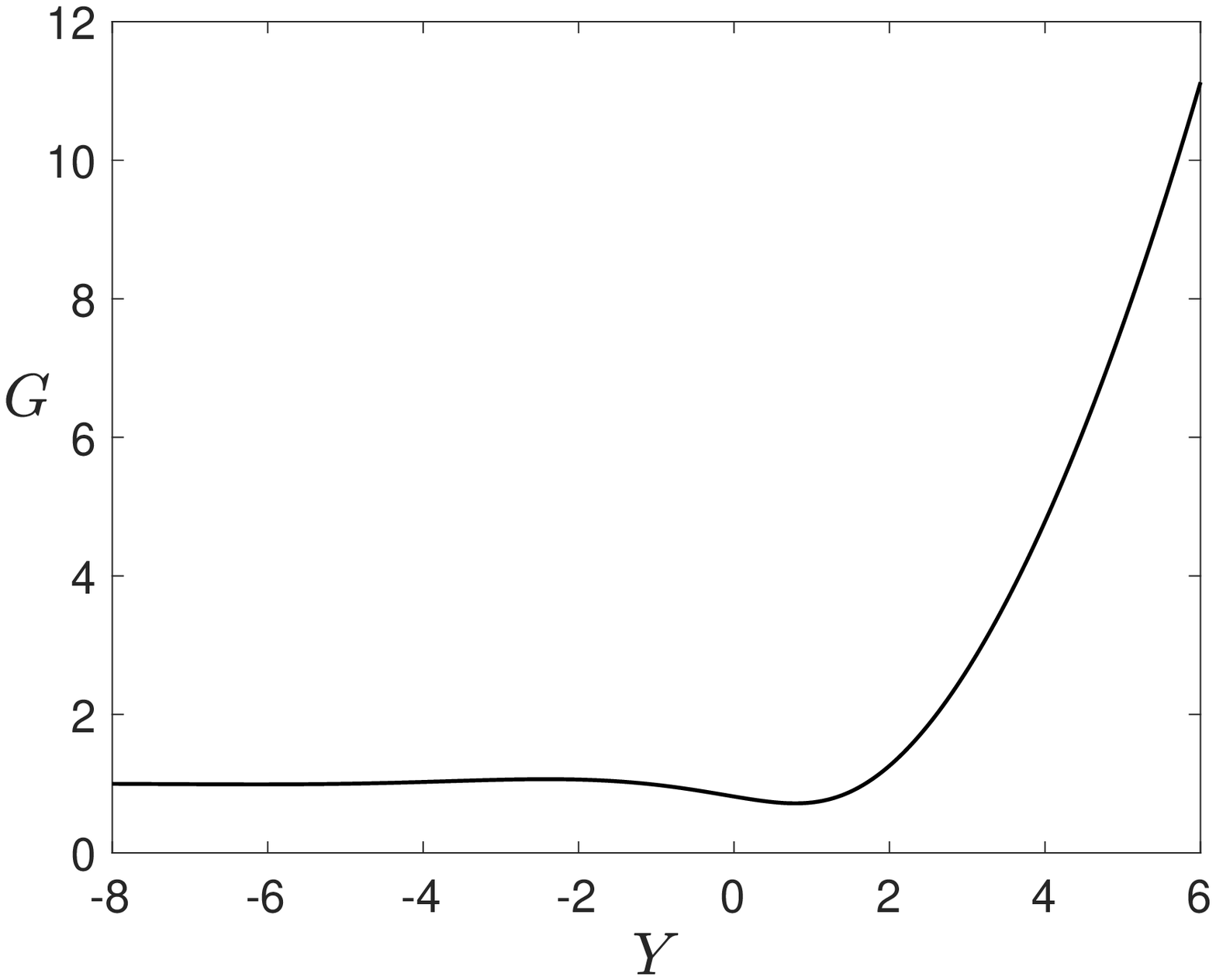}
\caption{}
\end{subfigure}
\caption{(a) Solution $F(y)$ to the problem (\ref{landau}), (\ref{landau_bc}). (b) Solution $G(Y)$ to the problem (\ref{app:Gprob}), (\ref{app:Gbc})}
\mylab{fig:Fproblem}
\end{figure}
%
In region $B_1$ we write $(z+\ell) = \delta^{1/2}\xi$, where $\xi=O(1)$, and expand by writing
$h = \delta H_0(\xi) + \cdots$. Substituting into (\ref{eq:app1}) and integrating once we obtain
\begin{eqnarray}\label{region_b1_eq}
H_0^3H_{0\xi\xi\xi} - c_0H_0 = Q_0.
\end{eqnarray}
Matching with regions $R_1$ and $R_2$ requires that
\bea \label{app:R1R2match}
H_0 \sim h_{p0} \,\,\, \mbox{as $\xi \to -\infty$}, \qquad
H_0 \sim a_0\xi^2 + h_1(-L) - \frac{c_0\Ca}{2a_0^2}\xi^{-1} \,\,\, \mbox{as $\xi \to \infty$},\quad
\eea
respectively. Here $h_{p0}$ is the scaled leading-order precursor film thickness to be determined.
If we rescale by writing $\xi = (h_{p0}/c_0^{1/3})\,\zeta$, $H_0(\xi) = h_{p0}F(\zeta)$, 
the problem takes the form \cite[see also][]{bretherton1961motion} 
\bea \label{landau}
F^3 F_{\zeta\zeta\zeta} - F + 1 = 0,
\eea
with
\bea \label{landau_bc}
F \sim 1 \quad \mbox{as $\zeta \to -\infty$}, \qquad
F \sim \mu_0 \zeta^2 + \mu_1\zeta + \mu_2  \quad \mbox{as $\zeta \to \infty$},
\eea
where $\mu_0 = a_0h_{p0}/c_0^{2/3}$, and $\mu_1$, $\mu_2$ are constants to be found. 
Useful insight is obtained by reformulating the problem as the first-order system $(u_1,u_2,u_3)_\zeta = (u_2,u_3,u_1^{-2} - u_1^{-3})$ with $(u_1,u_2,u_3)=(F,F_\zeta,F_{\zeta\zeta})$. 
It is straightforward to show that the fixed point at $(1,0,0)$ has a one-dimensional unstable manifold and a two-dimensional stable manifold. Thus, if it exists, the solution that fulfils the boundary conditions \eqref{landau_bc} is unique up to a translation in $\zeta$.  This freedom allows us to fix $\mu_1=0$ so as to satisfy \eqref{app:R1R2match} and match with region $R_2$. Solving numerically, we determine that $\mu_0 = 0.3215$ and $\mu_2 = 2.8996$ in exact agreement with the values calculated by
\cite{yu2013velocity}.
%
%
%
The numerical solution is shown in figure \ref{fig:Fproblem}.

Similar scalings apply in region $B_2$. Writing $z-\ell = \delta^{1/2}\tilde \xi$ and 
$h = \delta \tilde H_0(\tilde \xi) + \cdots$, the leading-order equation is found to be identical to \eqref{region_b1_eq}
but with tildes over all of the symbols except $c_0$.
Matching to regions $R_2$ and $R_3$ requires that
\bea \label{app:R2R3match}
\tilde H_0 \sim a_0 \tilde \xi^2 + h_1(\ell)  - \frac{c_0\Ca}{2a_0^2}\tilde \xi^{-1} \quad \mbox{as $\tilde\xi \to -\infty$},
\quad
\tilde H_0 \to h_{p0} \quad \mbox{as $\tilde \xi \to \infty$},
\eea
respectively. Rescaling so that $\tilde \xi = (h_{p0}/c_0^{1/3})\,Y$, $\tilde H_0(\tilde \xi) = h_{p0} G(Y)$, we have
\bea \label{app:Gprob}
G^3G_{YYY} - G + 1 = 0,
\eea
with 
\bea \label{app:Gbc}
G \sim \mu_0 Y^2 + \nu_1 Y + \nu_2 + \cdots \quad \mbox{as $Y \to -\infty$},
\qquad
G \sim 1 \quad \mbox{as $Y \to \infty$},
\eea
for constants $\nu_1$, $\nu_2$. The translational invariance with respect to $Y$ affords the freedom to set $\nu_1=0$ as required by the match with region $R_2$ via \eqref{app:R2R3match}.

\sloppy
Recasting as a first-order system it is readily seen that the fixed point at $(G,G_Y,G_{YY})=(1,0,0)$ has a two-dimensional stable manifold indicating that \eqref{app:Gprob}--\eqref{app:Gbc} has a one-parameter family of solutions for $G(Y)$.
The numerical solution, for which $\mu_0$ is set to the value computed above in region $B_1$, determines that
$\nu_2 = -0.8453$. This is in exact agreement with the value given by \cite{yu2013velocity}.

Using the above results we have that $h_1(-\ell) = \mu_2 h_{p0}$ and $h_1(\ell) = \nu_2 h_{p0}$. Then using \eqref{app:h1Leq} we obtain the value of the scaled leading-order the precursor film thickness
\bea\label{hp0asym}
h_{p0} = \frac{2h_{1P}(\ell)}{\nu_2-\mu_2} = 1.80,
\eea
and the leading-order wave speed coefficient
\bea \label{c0asym}
c_0 = \left( \frac{a_0h_{p0}}{\mu_0}\right)^{3/2} = 2.83.
\eea


\section{Travelling-wave computational method for Stokes flow}\label{sec:stokesflow}

The thin-film models are formally restricted to small Bond number and the requirement that $\cot \beta = O(\epsilon^{-1})$, as was discussed in 
\S\ref{sec:long}. The latter condition means that the thin-film model breaks down when $\beta$ is sufficiently close to $\pi$ and the inverted wall is 
almost horizontal. 
In \S\ref{sec:numerics}, we shall present results based on the thin-film models for angles $\beta$ that are very close to $\pi$, and also for large amplitude surface deformations. To 
allow us to corroborate these calculations we herein extend the discussion to parameter regimes beyond the range of validity of the thin-film models, 
and present a numerical boundary-integral method for computing travelling waves in Stokes flow (see, for example, \cite{pozred92} for a discussion of the theoretical formulation for such methods). 

We work in a frame of reference that is travelling with the wave at speed $c$. Here and henceforth all variables have been made dimensionless according to the scales mentioned in \S\ref{sec:problem}. We decompose the velocity field in the fluid by writing 
\bea
\bu = (\zh{U}(y) - c\bm{i}) + \bul(\zh{x},t),
\eea
where $\zh{U} = (u,v)$ is the dimensionless form of the Nusselt solution \eqref{nusselt}, $\zh{i}$ is the unit vector in the $x$ direction, and 
$\bul$ is the disturbance field that vanishes at the wall and which is to be found. The fluid stress $\zh{f}$ at a point $(x,y)$ on the free surface is similarly split up so that $\zh{f} = \zh{F} + \overline{\zh{f}}$, where $\zh{F}$ is the dimensionless Nusselt stress given by
%
%
\bea \label{basicF}
\zh{F} = -2\cos \beta\, (\sin^{-1/3}\beta - y) \zh{n} +
2\sin \beta\,  (\sin^{-1/3}\beta - y)
\begin{pmatrix}
0 & 1\\
1 & 0
\end{pmatrix}
\zh{n},
\eea
and $\zh{n}$ is the unit normal vector at the surface pointing into the fluid. 

The disturbance velocity and traction fields satisfy the Fredholm integral equation of the second kind for the disturbance velocity, 
\bea \label{bie}
\hspace{-0.4cm}2\upi \, \ul_j(\bx_0) = -S_j(\bx_0) + D_j(\bx_0),
\eea
for $j=1,2$ at a point $\bx_0=({x}_0,{y}_0)$ that is located on the free surface,
labelled $\mathscr{C}$, where
\bea
\!\!\!\!\!\!
S_j(\bx_0) = \int_{\mathscr{C}} G_{ij}(\bx,\bx_0)\fl_i(\bx)\,\dd s(\bx), \quad
D_j(\bx_0) = \int_{\mathscr{C}}^{\mathrm{p.v.}} \ul_i(\bx)T_{ijk}(\bx,\bx_0)n_k(\bx)\,\dd s(\bx)\,\,
\label{bie2}
\eea
are the single-layer and double-layer potential, respectively, and where 
p.v. denotes the principal value. In \eqref{bie2} $s$ is arc length along the free surface $\mathscr{C}$, and
$G_{ij}(\bx,\bx_0)$ and $T_{ijk}(\bx,\bx_0)$ are suitable choices for the Green's function and the stress tensor, respectively, for singularly-forced Stokes flow. 
In the travelling frame the kinematic condition requires that normal component of velocity on the free surface vanishes, so that $\bu\cdot \zh{n}=0$, and therefore
\bea \label{kincon2}
\bul = u_t\zh{t} -\zh{U} +  c\zh{i}
\eea
on $\mathscr{C}$, where $u_t=\bu\cdot \zh{t}$ is the {\it a priori} unknown tangential component of the total fluid velocity at the free surface, and $\zh{t}$ is the unit tangent pointing in the direction of increasing arclength.

The dynamic stress conditions \eqref{grapes} demand that $\zh{t}\cdot \zh{f} = 0$ and
$\zh{n}\cdot \zh{f} = \Ca^{-1} \kappa$ on $\mathscr{C}$, and hence that
\begin{eqnarray}
\label{grapes2}
\overline{\zh{f}} = \Ca^{-1} \kappa \zh{n} - \zh{F}
\end{eqnarray}
on $\mathscr{C}$, where $\zh{F}$ was given in \eqref{basicF}. Here the curvature $\kappa = -\bm{n}\cdot \bm{t}_s$ consistent with the definition made in \S\ref{sec:problem}.

The integral equation \eqref{bie} together with the kinematic condition \eqref{kincon2} and the dynamic stress conditions \eqref{grapes2} must be solved numerically. 
We work on a computational domain with periodic boundary conditions, and we
use the periodic Green's function $\zh{G}^{\mbox{\tiny{PW}}}$ that has the property $\zh{G}^{\mbox{\tiny{PW}}}(\bx,\bx_0)=\zh{0}$ when $\bx$ lies on the wall at $ y=0$ \cite[see, for example,][]{pozred92}
\bea \label{greenSP}
\zh{G}^{\mbox{\tiny{PW}}}(\bx,\bx_0) = \zh{G}^{\mbox{\tiny{P}}}(\zh{\hat x}) - \zh{G}^{\mbox{\tiny{P}}}(\zh{\hat X}) + 2{y}_0^2\zh{G}^{\mbox{\tiny{DP}}}(\zh{\hat X})
-2{y}_0\zh{G}^{\mbox{\tiny{SDP}}}(\zh{\hat X}),
\eea
where $\zh{\hat x}=\bx-\bx_0$ and $\zh{X}=\bx-\bx_0'$ with $\bx_0'=({x}_0,-{y}_0)$.
In  \eqref{greenSP} the Green's function
\bea \label{GP}
\zh{G}^{\mbox{\tiny{P}}}({\zh{x}}) = 
\begin{pmatrix}
1 - A -  yA_{ y} &  yA_{ x} \\
 yA_{ x} &  yA_{ y}-A
\end{pmatrix},
\eea
where
\bea
A(\bx) = \frac{1}{2}\log\left(\cosh (2\pi  y/L) - \cos (2\pi  x/L)\right) + \frac{1}{2}\log 2
\eea
corresponds to a periodic array of Stokeslets, and $\zh{G}^{\mbox{\tiny{DP}}}$, $\zh{G}^{\mbox{\tiny{SDP}}}$ are the periodic potential dipole and periodic Stokeslet doublet, 
respectively, both given in closed form in \cite{pozred92}. 
An alternative form for the Green's function and stress 
tensor, derived using a complex variable approach, was recently provided by 
\cite{crowdy2019analytical}.

We compute the solution to the integral equation \eqref{bie} with spectral accuracy by adapting the protocol proposed by \cite{veerapaneni2009boundary} for the motion of vesicles in a Stokes flow. We describe a point $\bx(\alpha)=( x(\alpha), y(\alpha))$ on the free surface, and the tangential surface velocity, in the form
\bea \label{xfft}
\bx(\alpha) = \frac{\alpha L}{\pi}\zh{e}_x + \sum_{n=-N}^{N} \zh{\hat x}_n\ee^{\ri n \alpha},
\qquad
u_t(\alpha) = \sum_{n=-N}^{N} \hat u_n\ee^{\ri n \alpha},
\eea
where $2L$ is the domain size, $\alpha \in [0,2\pi)$ is a parameter, $\zh{\hat x}_n$ and $\hat u_n$ are complex coefficients to be found, and $N$ is a specified truncation level. Since $\bx$ and $u_t$ are real, $\zh{\hat x}_n=\zh{\hat x}_{-n}^*$ and $\hat u_n = \hat u_{-n}^*$, where the asterisk denotes the complex conjugate. 

Inserting \eqref{kincon2} and \eqref{grapes2} into \eqref{bie} we enforce the resulting integral equation at a set of $2N+1$ equally-spaced collocation points 
$\bx_0=\bx(\alpha_0)$, where $\alpha_0 \in \{2(k-1)\pi/(2N+1): k=1,\ldots,2N+1\}$. The necessary $\alpha$ derivatives 
are computed at each collocation point using a fast Fourier transform to yield numerical approximations for the unit normal and tangent vectors, and for the free-surface curvature. The single-layer potential $S_j({\zh{x}_0})$ is weakly singular as $\bx\to \bx_0$, and in particular the Green's function and stress tensor 
tend toward the two-dimensional Stokeslet
\bea \label{stokeslet}
G^{\tiny{\mbox{ST}}}_{ij} = -\delta_{ij}\log {r} + \frac{\hat x_i\hat x_j}{{r}^2}, \qquad T^{\tiny{\mbox{ST}}}_{ijk} = -4\frac{\hat x_i\hat x_j\hat x_k}{{r}^4},
\eea
where $ r=|\bx-\bx_0|$.
The integrand of the single-layer potential is therefore logarithmically singular in this limit and, following \cite{veerapaneni2009boundary}, we 
calculate it numerically with spectral accuracy using the hybrid Gauss-trapezoidal 
quadrature formula of \cite{alpert1999hybrid}. This hybrid formula assumes the presence of a logarithmic singularity at the lower integration limit and applies the usual
trapezoidal rule in the centre of the integration range with a weighted Gaussian quadrature at each 
end.
Specifically, it supplies the approximation
\bea \label{quad}
\hspace{-0.25in} \mathcal{S}_{ij}(\alpha_0) \approx \hs \sum_{k=1}^{N_L} w^{L}_k S_{ij}(v^{L}_k \hs,\alpha_0)
+ \hs \sum_{k=0}^{N_T} S_{ij}(a\hs + k\hs,\alpha_0)
+ \hs \sum_{k=1}^{N_R} w^{R}_k S_{ij}(v^{R}_k \hs,\alpha_0),
\eea
where the $w_k$ and $v_k$ with $L$/$R$ superscripts are $N_L$, $N_R$ weights and nodes at the left-hand and right-hand ends respectively; $N_T$ is the number of trapezium rule points with spacing $\hs=1/(N_T+a+b-1)$, where
$a$ and $b$ are parameters that are related to the convergence properties of the quadrature. In our calculations we took $a=10$ and $b=7$ with $N_L=15$ and $N_R=8$ to achieve convergence on the order $O(\hs^{16}\log \hs)$  \cite[see Tables 6 and 8 of][where numerical values for the weights and nodes are given]{alpert1999hybrid}. As suggested by  
\cite{veerapaneni2009boundary}, we split the integration of the single-layer potential up into the ranges $[0,\alpha_0]$ and $[\alpha_0,2\pi]$ and apply the quadrature rule \eqref{quad} appropriately over each range.
For the double layer potential we apply the regular quadrature rule of \cite{alpert1999hybrid} which is obtained by setting $a=b$ and using the same weights and nodes at both ends in \eqref{quad}; in this case we took $a=7$ to achieve order $O(\hs^{16})$ convergence \cite[see Table 6 of][]{alpert1999hybrid}. 

Next, we express the boundary integral equation \eqref{bie} in the form 
{\color{black} and we enforce the conditions
\bea \label{Rproj}
\zh{\mathcal{R}}\cdot \zh{n} = 0, \qquad \zh{\mathcal{R}}\cdot \zh{t} = 0, \qquad 
\eea
at the $2N+1$ collocation points ${\zh{x}}_0={\zh{x}}(\alpha_0)$ defined above, yielding $4N+2$ algebraic equations}. 
A further $2N$ equations follow by demanding
\bea
\int_0^{2\pi} |{\zh{x}}_\alpha|\,\ee^{\ri n\alpha}\,\dd \alpha = 0
\eea
for $n=\pm 1, \, \pm 2,\ldots, \pm N$,
so that $s_\alpha =  |{\zh{x}}_\alpha|$ is constant along $\mathscr{C}$ and the collocation points are equally spaced with respect to arc length along the free surface. 
One further equation arises by fixing the origin in $ x$, specifically by setting $ x(0)=-L$.
The final equation needed is supplied either by fixing the height of the film at one end of the domain in the case of a fixed flow rate calculation, setting $ y(0)=h_0$, or else by fixing the fluid volume as
\bea
\int_{-L}^L  y\,  \dd  x = \int_0^{2\pi}  y  x_{\alpha}\,\dd \alpha = V.
\eea
The translational invariance of the system is removed by fixing the free-surface maximum to lie in the middle of the domain, setting
\bea
y_\alpha(\pi) = 0.
\eea
For any $\bx_0$, conservation of mass implies that \cite[][]{pozred92}
\bea
\int_{\mathscr{C}} G^{\mbox{\tiny{PW}}}_{ij}(\bx,\bx_0)n_j(\bx)\,\dd s(\bx) = 0,
\eea
and so, referring to \eqref{bie}, the disturbance stress $\overline{\zh{f}}$ is determined to within an arbitrary constant multiple of $\bm{n}$. Consequently one equation can be removed from the projection $\zh{\mathcal{R}}\cdot \zh{n}=0$ in \eqref{Rproj} arbitrarily to obtain a system of $6N+4$ for the $6N+4$ unknowns comprising
the $6N+3$ Fourier {\color{black} coefficients $\{\zh{\hat x}_n,\hat u_n\}$}, and $c$. 
This system is solved using Newton's method, wherein 
at each stage we compute the Fourier representation \eqref{xfft} using a fast Fourier transform.

\section{Numerical results}\label{sec:numerics}

In this section we study the behaviour of steady pulse solutions in the two cases of fixed flow rate and fixed volume. Of particular interest is to follow the solution branch for a pulse as $\beta$ increases and the wall tends to become horizontal, a limit that we naturally associate, on the basis of physical intuition, with the onset of dripping.
\begin{table}
\centering
\begin{tabularx}{270pt}{| l| >{\centering\arraybackslash}X | >{\centering\arraybackslash}X | >{\centering\arraybackslash}X | >{\centering\arraybackslash}X | >{\centering\arraybackslash}X | }
\hline\\[-0.565cm]
& $\Ca$ & $V$ & $\Ca V$ & $q$ & Figure \\[-0.23cm]  \hline\\[-0.56cm]
Fixed volume & 0.005 & 400, \, 900 & 2.0,\,\, 4.5 & --- & \ref{fig:thinStokes_vol}, \,\,\,\, \ref{fig:YL-FCMcomp3} \\
& 0.3     & \,8, \,\,\,\,\, 15    & 2.4,\,\, 4.5 & --- & \ref{fig:YL-FCMcomp}, \,\, \ref{fig:YL-FCMcomp2}\\[-0.23cm]  \hline\\[-0.565cm]
Fixed flow rate & 0.005 & --- & --- & 2/3 & \ref{fig:thinStokes}\\
& 0.3     & --- & --- & 2/3 & \ref{fig:YL-FCM_fixed_q}
\\[-0.23cm]  \hline
\end{tabularx}
\caption{Summary of the calculations in \S\ref{sec:numerics}. Note that the turning point for the YL fixed volume solutions in figure~\ref{fig:YLsoln} occurs at $\Ca V = 2.60$.}
\label{table}
\end{table}

Working on a periodic domain $[-L,L]$ we compute travelling-wave solutions to the long-wave LCM and FCM equations introduced in \S\ref{sec:long} using a scheme based on Newton iterations and a Fourier pseudospectral representation of the spatial derivatives \cite[see][]{blyth2018two, lin2018continuation}, and to the equations of Stokes flow using the numerical method discussed in \S\ref{sec:stokesflow}. Travelling-wave solution branches emerge from the neutral stability point where the growth rate of small amplitude periodic waves vanishes. The relationship between $\beta$, $\Ca$ and $L$ at the neutral stability point is
\bea\label{critrel}
L = \pi(-2\Ca \cos\beta)^{-1/2}
\eea
 for both the LCM and the FCM \cite[see][]{blyth2018two}, and for Stokes flow (see Appendix~\ref{appendixstokesflat}).
For a chosen $\Ca$ we fix the domain size $2L$ and calculate the inclination angle close to the critical value determined from \eqref{critrel} to compute a small amplitude periodic wave. We then perform pseudo-arclength continuation in $\beta$ to study localised drop formation as the wall tends to become horizontal. 
%
\begin{figure}
%
%
\begin{center}
\begin{subfigure}{0.5\textwidth}
\includegraphics[width=2.6in]{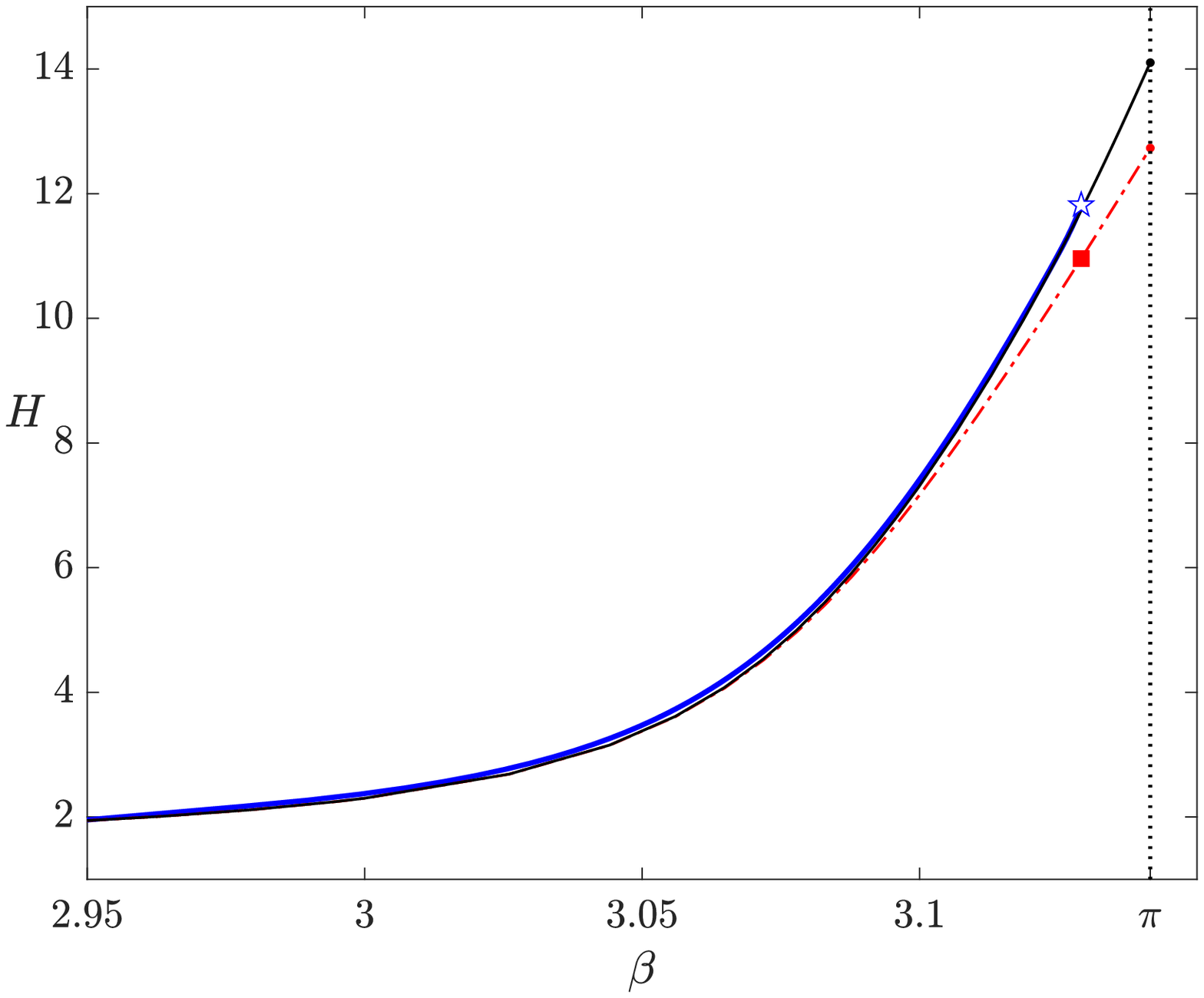}
\caption{}
\end{subfigure}
\end{center}
\vspace{0.075in}
\begin{subfigure}{0.5\textwidth}
\includegraphics[width=2.6in]{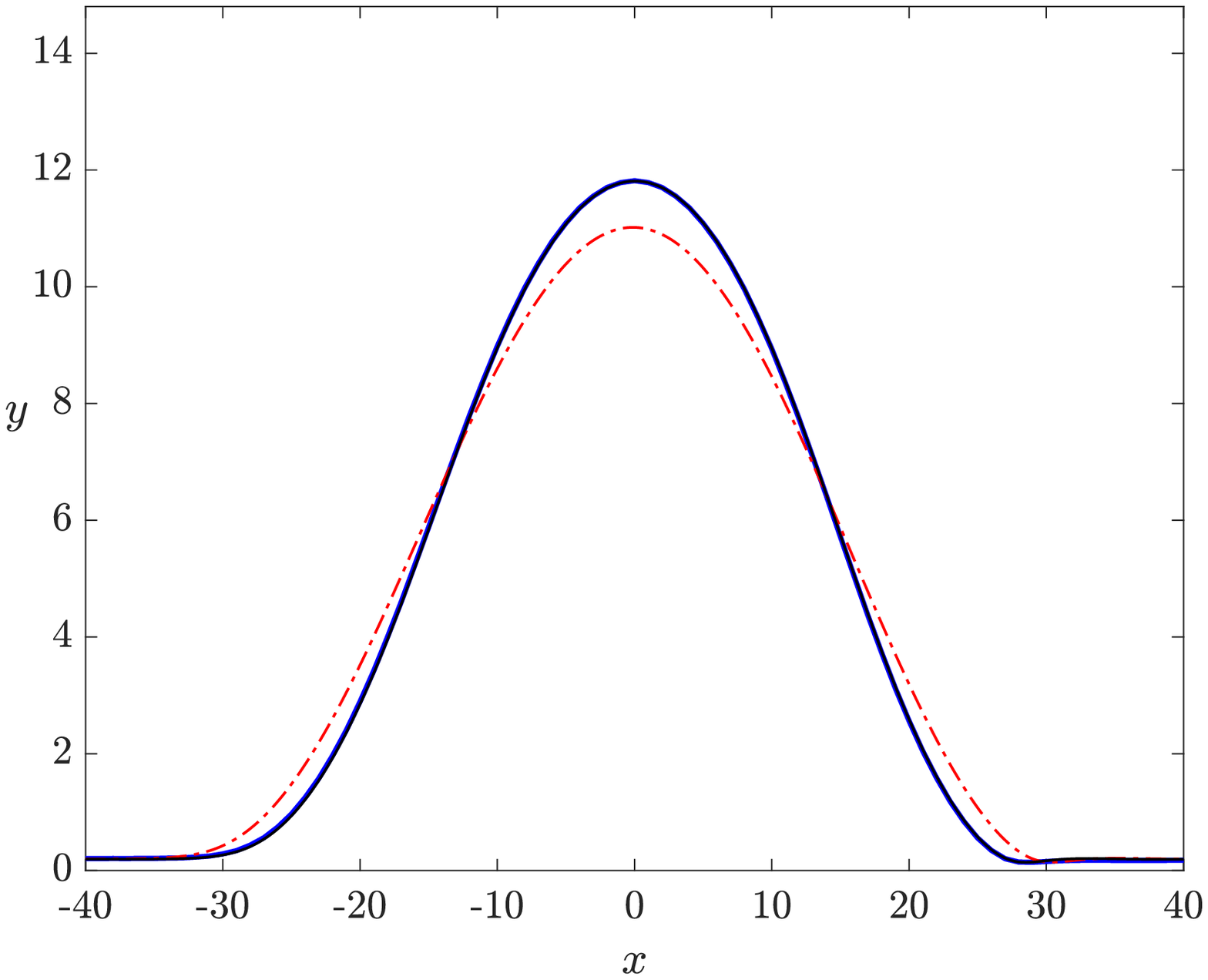}
\caption{}
\end{subfigure}
\begin{subfigure}{0.5\textwidth}
\includegraphics[width=2.6in]{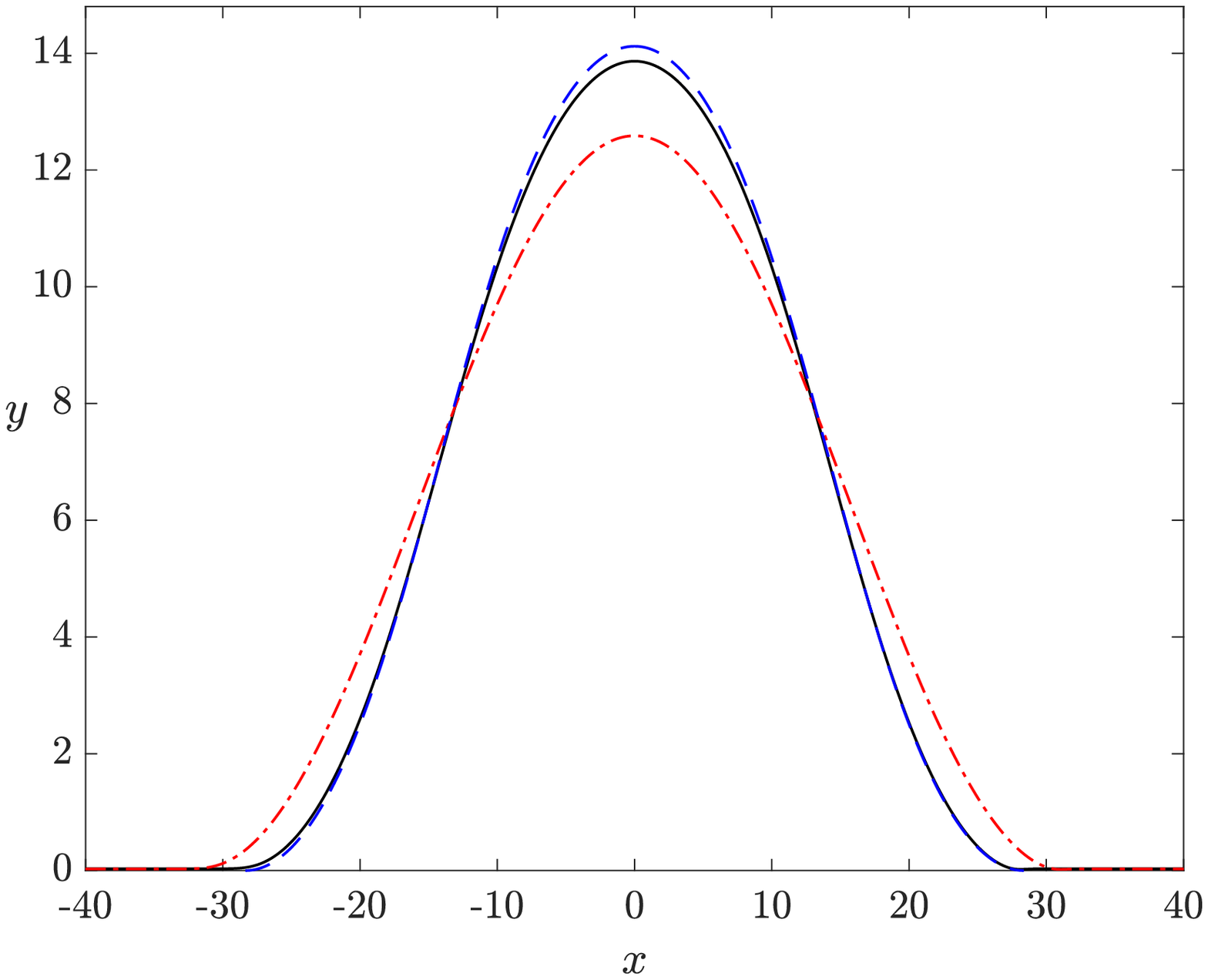}
\caption{}
\end{subfigure}
\caption{Fixed volume calculation for $\Ca=0.005$ and $V=400$ (so $\Ca V = 2.0$). Comparison between the boundary-integral calculation for Stokes flow, shown with a thick blue line, and the FCM and LCM, shown with solid black lines and dot-dashed red lines respectively. The computations were done on the domain $[-150,150]$. (a) Drop heights $H$ versus inclination angle $\beta$. (b, c) Drop profiles at $\beta=3.129$ and $3.14$ respectively,  corresponding to the pentagram and the filled square, and to the end points of the FCM and LCM curves in (a). 
%
%
The blue dashed line in panel (c) is the YL solution to equation \eqref{eq:YLnondim}.}
\mylab{fig:thinStokes_vol}
\end{figure}
%
%

In the following subsections we study the cases of fixed volume $V$ and fixed fixed flow rate $q$ separately. Throughout we shall refer to the maximum vertical distance between the wall and \cbl the film surface (measured in the $y$ direction, as depicted in Figure~\ref{fig1}) \cb as the drop height $H$, and we will use it as a measure of the solutions in the bifurcation diagrams we shall construct. In each case we consider computations at the small Bond number $\Ca=0.005$ in order to facilitate comparison between the long-wave models and Stokes flow, as well as the larger Bond number $\Ca=0.3$ in order to study the dynamics beyond the range of validity of the long-wave models. In the case of fixed volume, for each Bond number we select a couple of cases wherein $\Ca V$ is before and after the turning point in figure~\ref{fig:YLsoln}. Specifically we choose $V=400$ and $V=900$ for the case $\Ca=0.005$; and we choose $V=8$ and $V=15$ for the case $\Ca=0.3$ (see Table~\ref{table}). For fixed flow rate, for each of the two Bond numbers we set $q=2/3$ corresponding to a film of unit dimensionless thickness on a vertical wall according to \eqref{nondimh}.

\subsection{Fixed volume}

First we discuss the case of a small Bond number, taking $\Ca=0.005$, for which we expect to find good agreement between the predictions from the thin film models and the Stokes flow computations. After this we will examine the larger Bond number case $\Ca=0.3$ for which we expect to see significant differences.
With reference to table~\ref{table}, for $\Ca=0.005$ we consider the two volumes $V=400$ and $V=900$, and for
$\Ca=0.3$ we consider the two volumes $V=8$ and $V=15$. In both cases these values are chosen so that $\Ca V$ lies either side of the turning point at $\Ca V=2.60$ in figure~\ref{fig:YLsoln} which shows the bifurcation diagram for YL solutions on a perfectly inverted plate. In each set of results we show bifurcation diagrams of the drop height $H$ versus the inclination angle $\beta$ as well as the drop profiles at certain inclination angles. As is indicated in the figure captions, computations for the LCM are shown with red dot-dashed lines, those for the FCM are shown with thin black solid lines, and those for Stokes flow with thick blue solid lines. 
%
%
%
%
\begin{figure}
\begin{subfigure}{0.5\textwidth}
\includegraphics[width=2.6in]{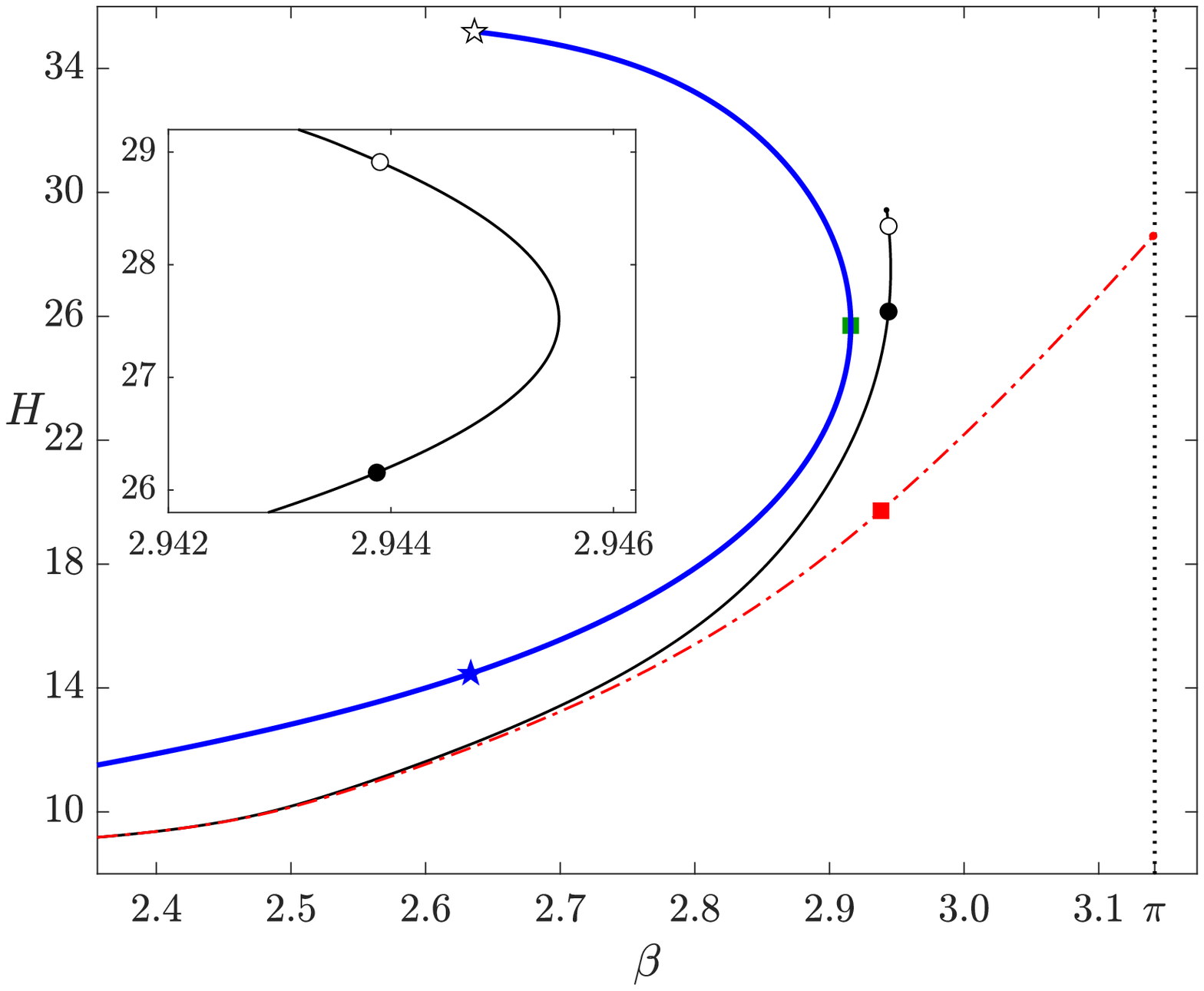}\hspace{0.025in}
\caption{}
\end{subfigure}
\begin{subfigure}{0.5\textwidth}
\includegraphics[width=2.6in]{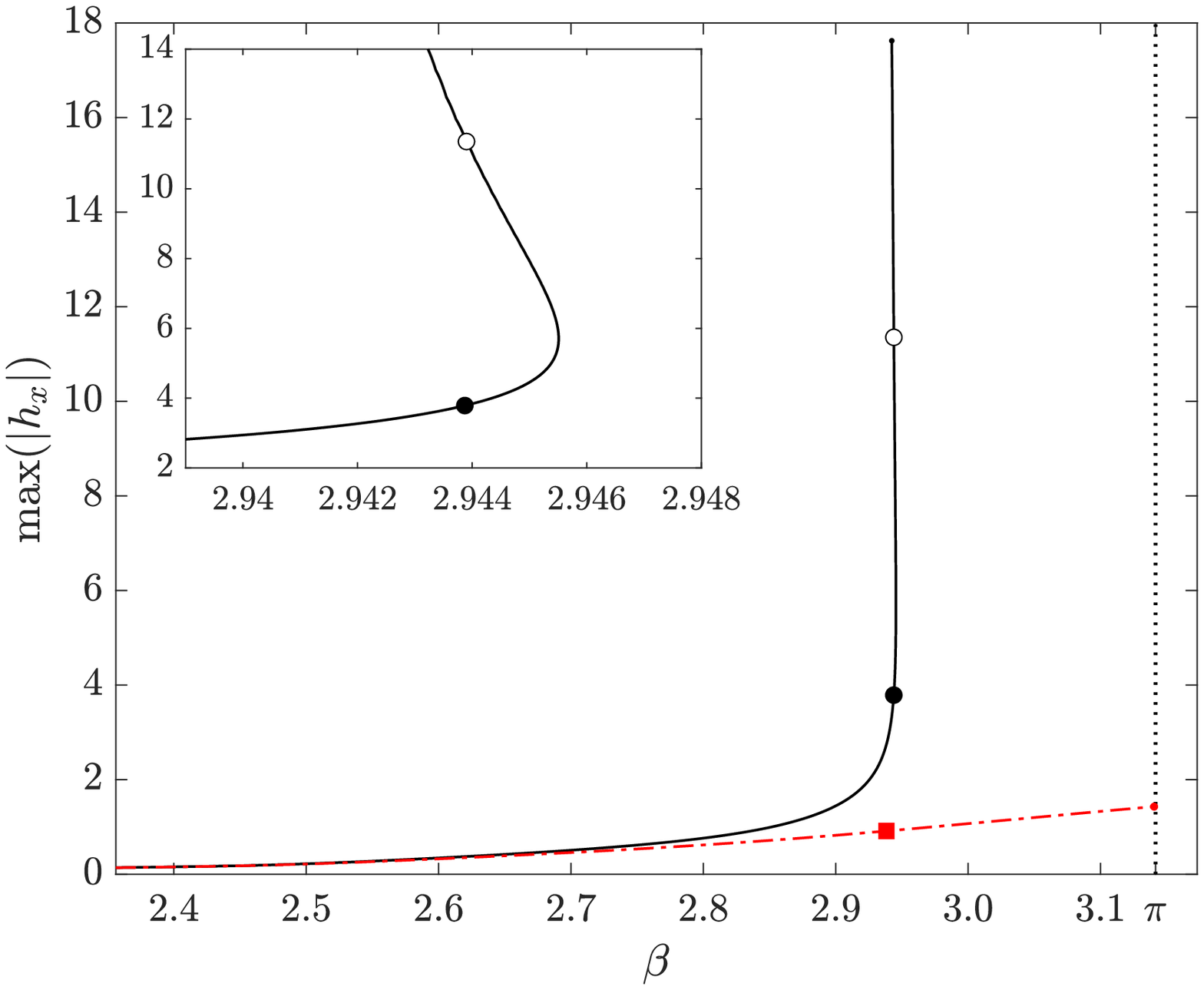}
\caption{}
\end{subfigure}
\begin{subfigure}{0.5\textwidth}
\includegraphics[width=2.6in]{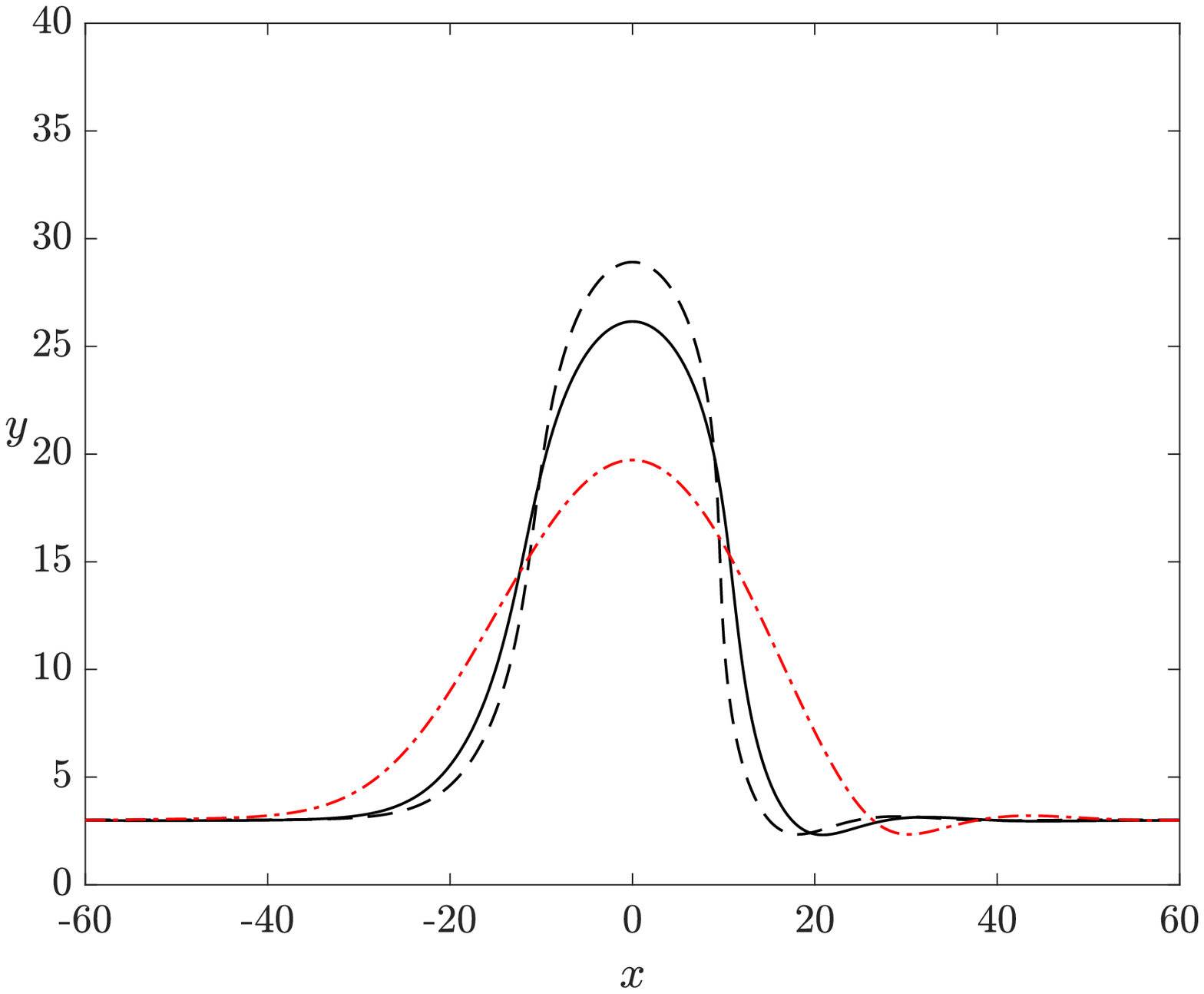}
\caption{}
\end{subfigure}
\begin{subfigure}{0.5\textwidth}
\includegraphics[width=2.6in]{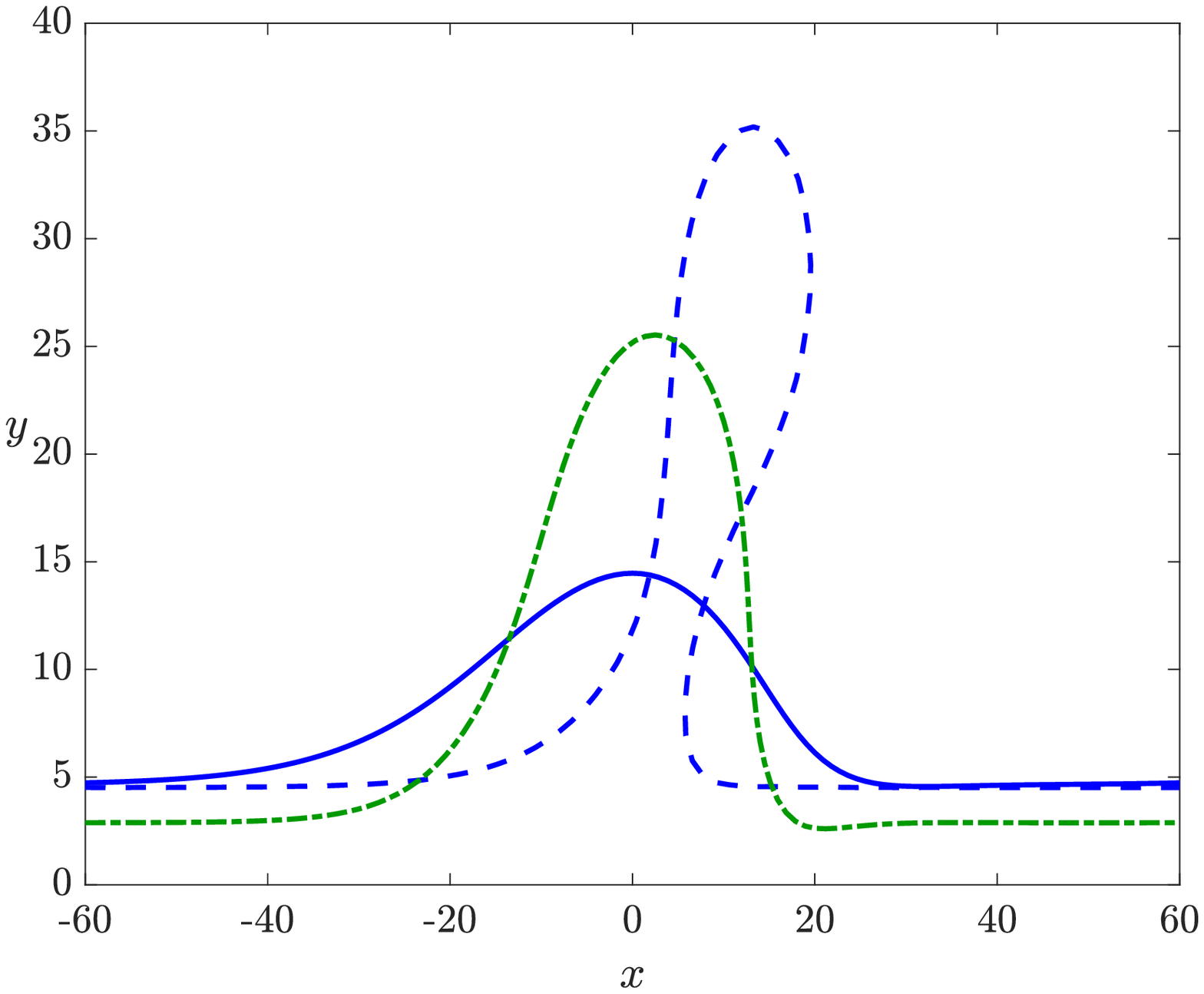}
\caption{}
\end{subfigure}
\caption{Fixed volume calculation with $\Ca=0.005$ and $V=900$ (so $\Ca V = 4.5$). Thin-film calculation for the FCM \eqref{eq:TFE2_full}, shown with solid lines, the LCM \eqref{eq:TFE1}, shown with dot-dashed lines, and the Stokes calculation, shown with thick solid lines, all computed on the domain $[-60,60]$. (a) Drop heights $H$ versus inclination angle $\beta$. (b) Maximum of the absolute value of the drop slope versus inclination angle $\beta$. (c) Drop profiles at $\beta=2.9439$ corresponding to the filled and empty circles (solid and dashed lines, respectively, for the FCM) and the filled square (dot-dashed line for the LCM) in the diagrams (a) and (b). (d) Drop profiles for Stokes flow at the filled and empty star symbols corresponding to $\beta=2.6335$ (solid and dashed lines, respectively) and at the filled square at $\beta=2.9158$ (dot-dashed line) corresponding to the turning point.}
\mylab{fig:YL-FCMcomp3}
\end{figure}
%
%

%
\begin{figure}
\begin{center}
\begin{subfigure}{0.5\textwidth}
\includegraphics[width=2.6in]{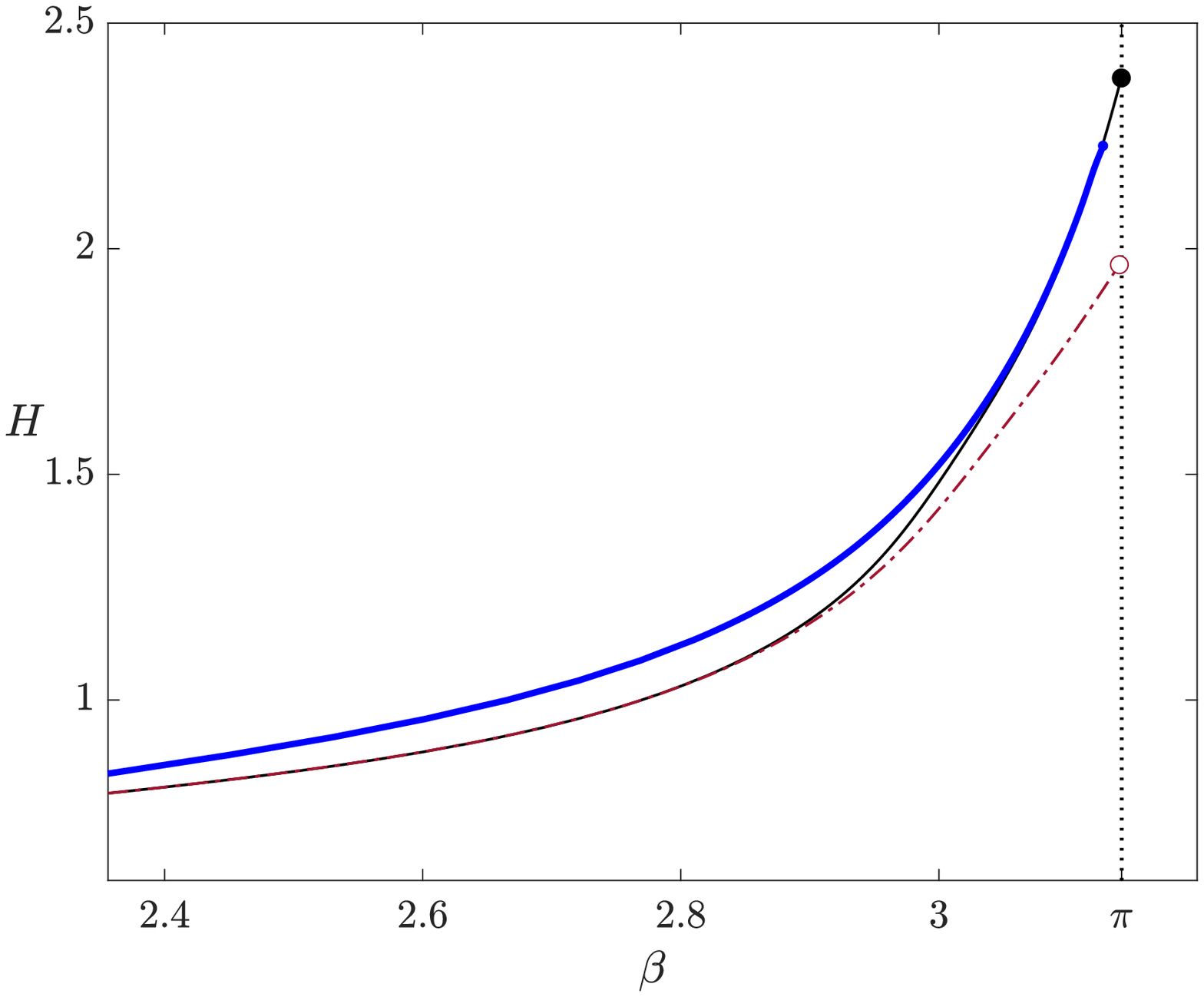}
\caption{}
\end{subfigure}
\end{center}
\vspace{0.075in}
\begin{subfigure}{0.5\textwidth}
\includegraphics[width=2.6in]{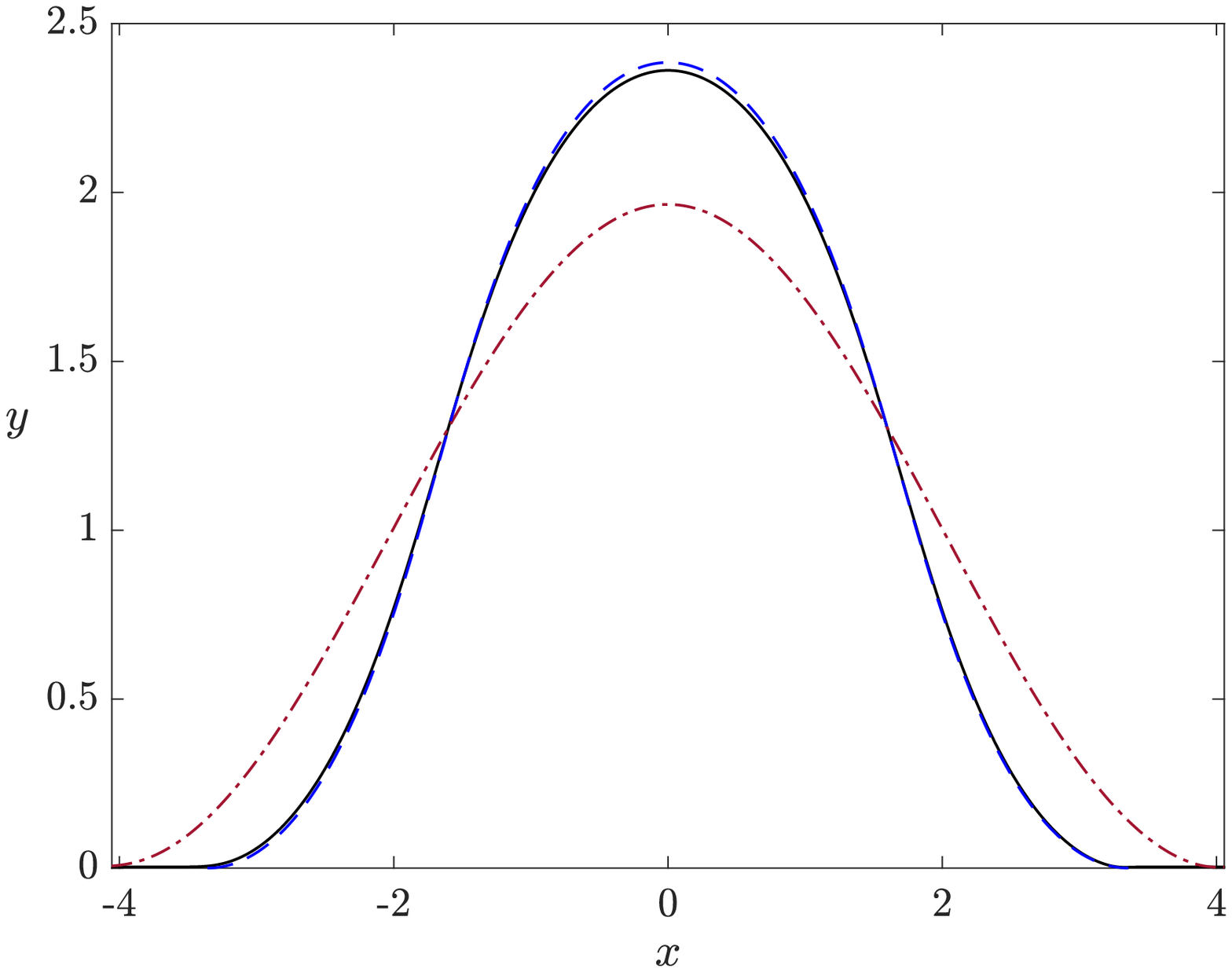}
\caption{}
\end{subfigure}
\begin{subfigure}{0.5\textwidth}
\includegraphics[width=2.6in]{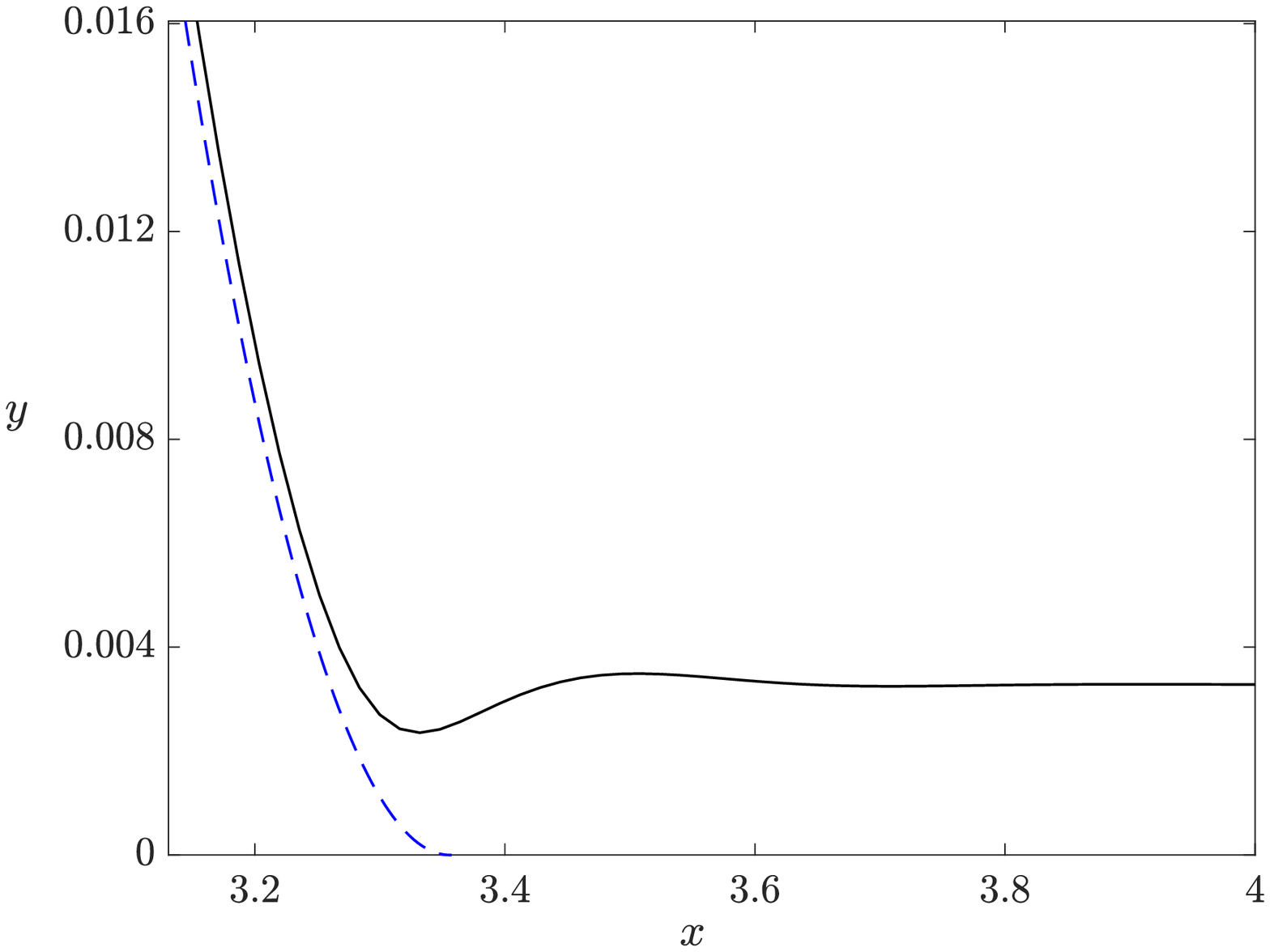}
\caption{}
\end{subfigure}
\vspace{-0.15in}
\caption{Fixed volume calculation with $\Ca=0.3$ and $V=8$ (so $\Ca V = 2.4$). Thin-film calculation for the FCM \eqref{eq:TFE2_full}, shown with solid lines, and the LCM \eqref{eq:TFE1}, shown with dot-dashed lines, and the Stokes calculation shown with a thick solid line, all on the domain $[-6,6]$. (a) Drop heights $H$ versus inclination angle $\beta$. (b) Drop profiles at $\beta=3.1398$ (shown with filled and empty circles in panel a)
including a comparison with the YL equation \eqref{eq:YLnondim} for $\beta=\pi$ (for which $L=3.36$ is found), shown with a dashed line. (c) Close-up of the drop profiles for the FCM and YL models.} 
\mylab{fig:YL-FCMcomp}
\vspace{0.175in}
%
\begin{subfigure}{0.5\textwidth}
\includegraphics[width=2.26in]{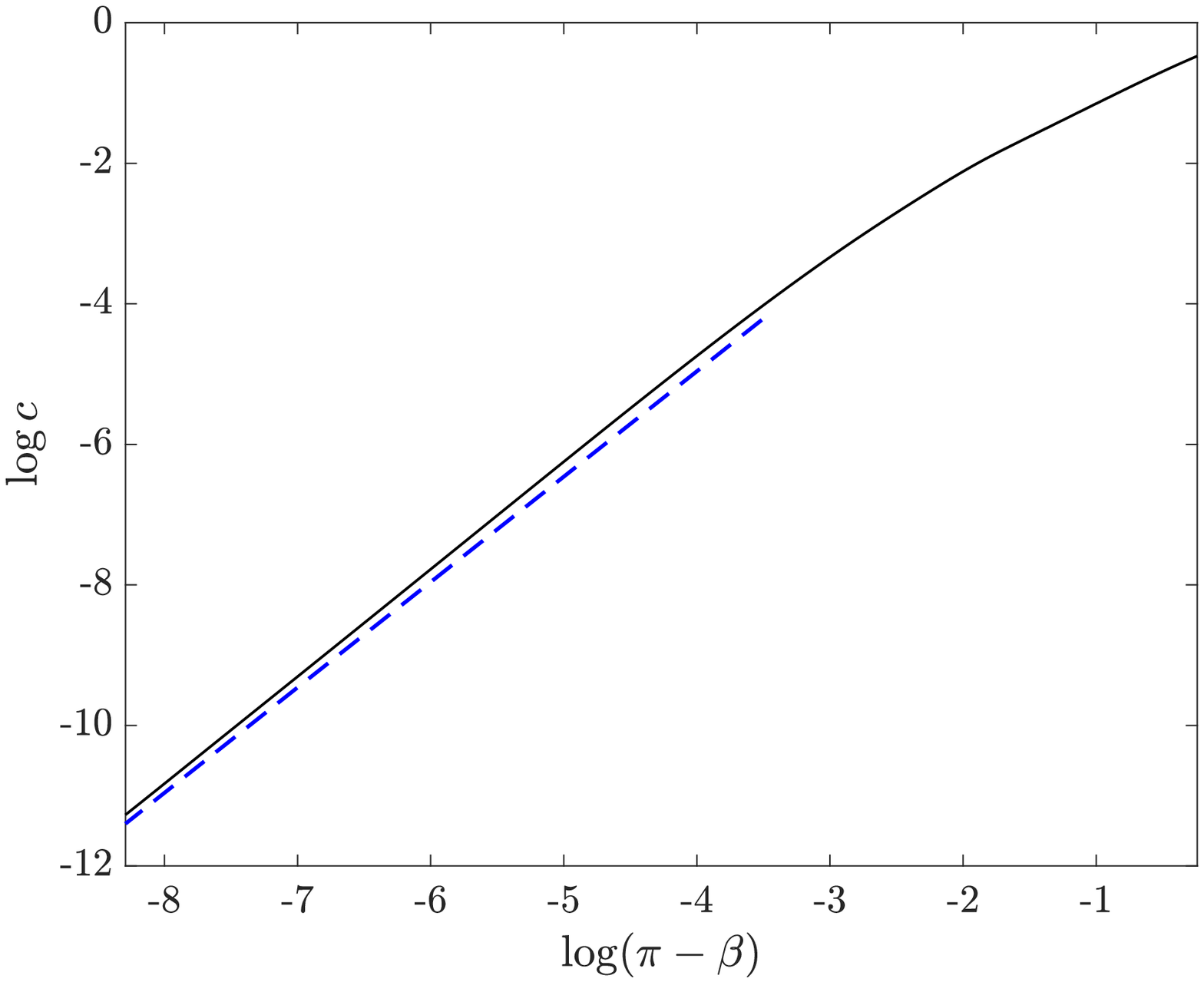}
\caption{}
\end{subfigure}
\begin{subfigure}{0.5\textwidth}
\includegraphics[width=2.26in]{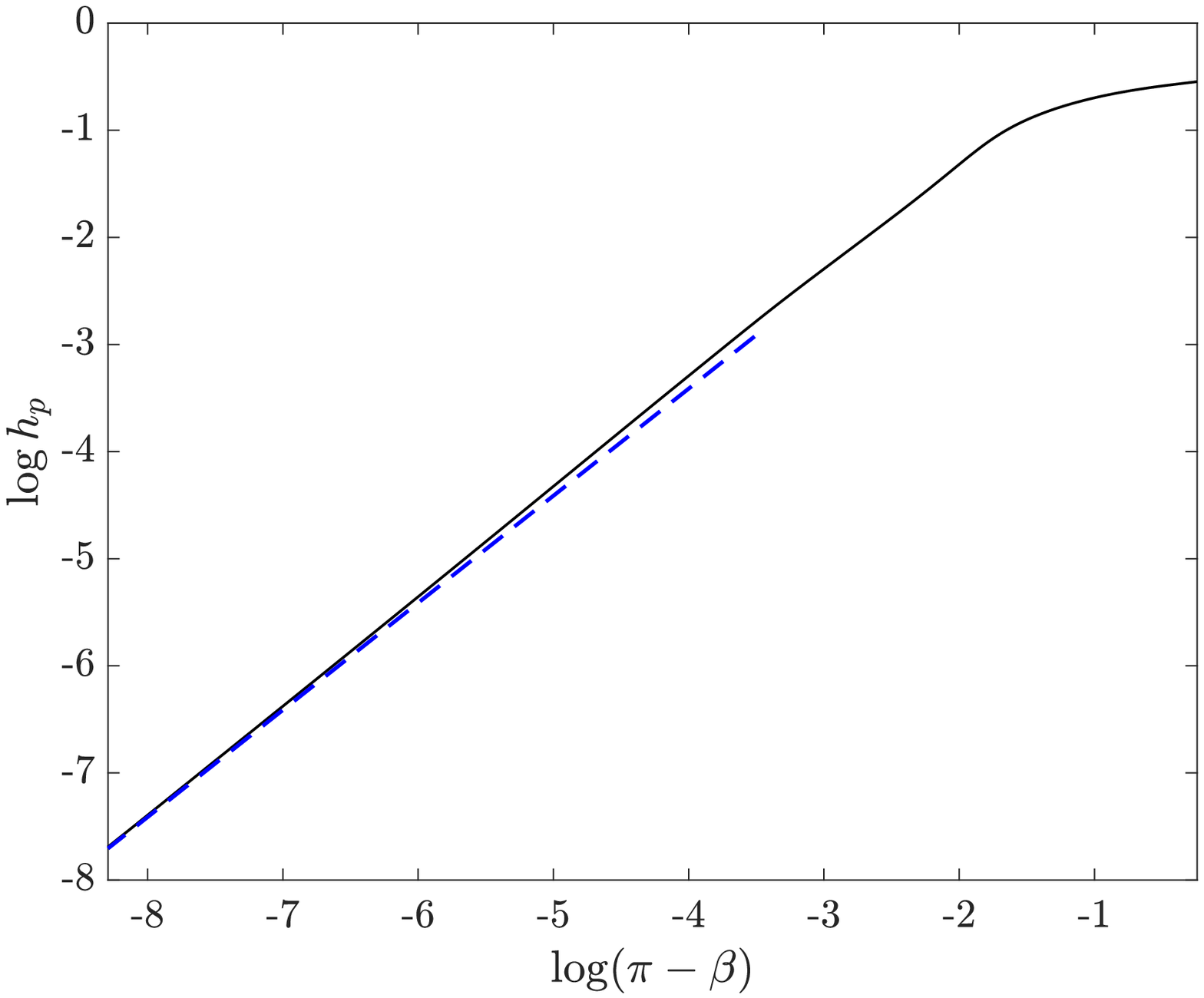}
\caption{}
\end{subfigure}
\vspace{-0.15in}
\caption{Behaviour of the wave speed $c$ and the precursor film thickness $h_{p}$ near to $\beta=\pi$ for the calculation in figure \ref{fig:YL-FCMcomp}. The dashed lines correspond to the asymptotic estimates $c = (\pi-\beta)^{3/2}c_0$, with $c_0=2.83$ according to \eqref{c0asym}, and $h_p = (\pi-\beta) h_{p0}$, with $h_{p0}=1.80$ according to \eqref{hp0asym}.}
\mylab{fig:chasym}
\end{figure}
%
%
In the small Bond number case with $\Ca V = 2.0$ shown in figure~\ref{fig:thinStokes_vol}, there exists a YL solution for the fully inverted plate at $\beta=\pi$. We see that the bifurcation curves in panel (a) all continue to $\beta=\pi$ and that those for the FCM and for Stokes flow both approach the static YL solution. Confirmation of this is provided in panel (c) of the same figure, where the drop profile for the FCM at $\beta=3.14$ is very close to the YL profile. The LCM approaches a pure cosine solution to~\eqref{YLlin}, with support $(2/\Ca)^{1/2}\pi \approx 62.8$, as expected. Computational difficulties obstruct the continuation of the Stokes flow solution beyond the value $\beta\approx 3.129$, where the corresponding bifurcation curve in the figure terminates. 
Nevertheless the drop profiles shown at $\beta=3.129$ in panel (b) confirm the excellent agreement between the FCM and Stokes flow computations.

%
%
%
%
%
\begin{figure}
\begin{subfigure}{0.5\textwidth}
\includegraphics[width=2.6in]{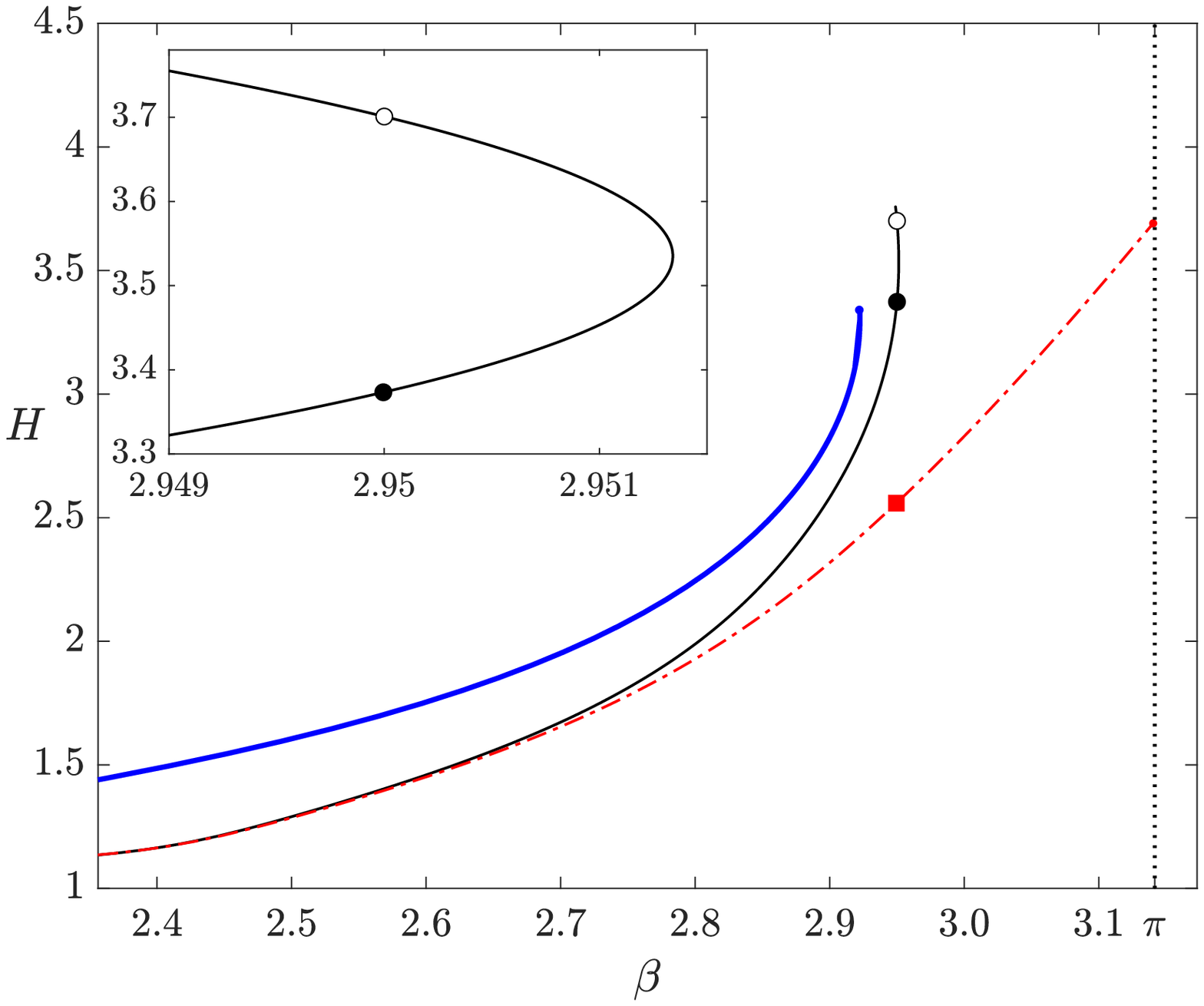}\hspace{0.025in}
\caption{}
\end{subfigure}
\begin{subfigure}{0.5\textwidth}
\includegraphics[width=2.6in]{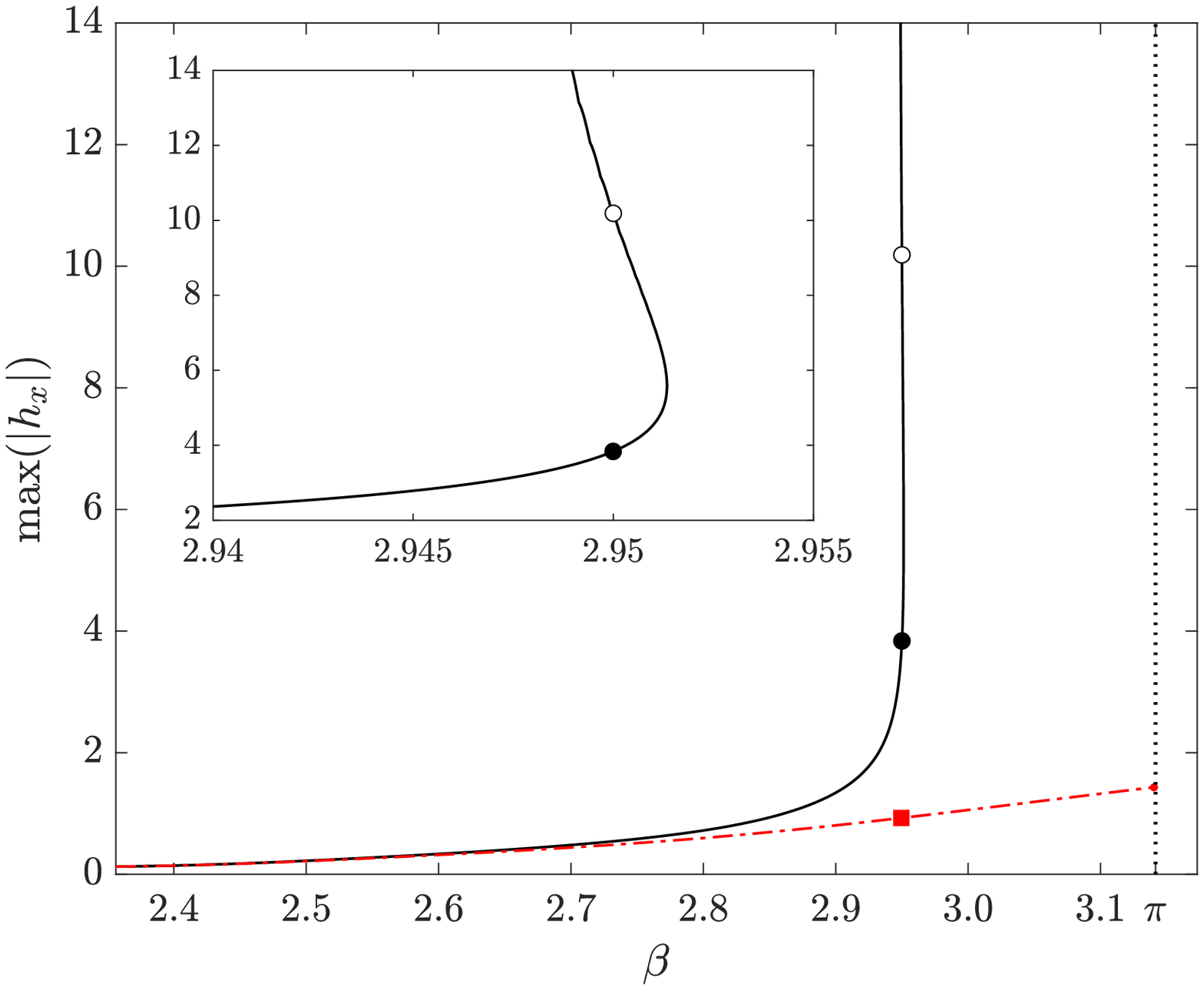}
\caption{}
\end{subfigure}
\\[0.075in]
\begin{center}
\begin{subfigure}{0.5\textwidth}
\includegraphics[width=2.6in]{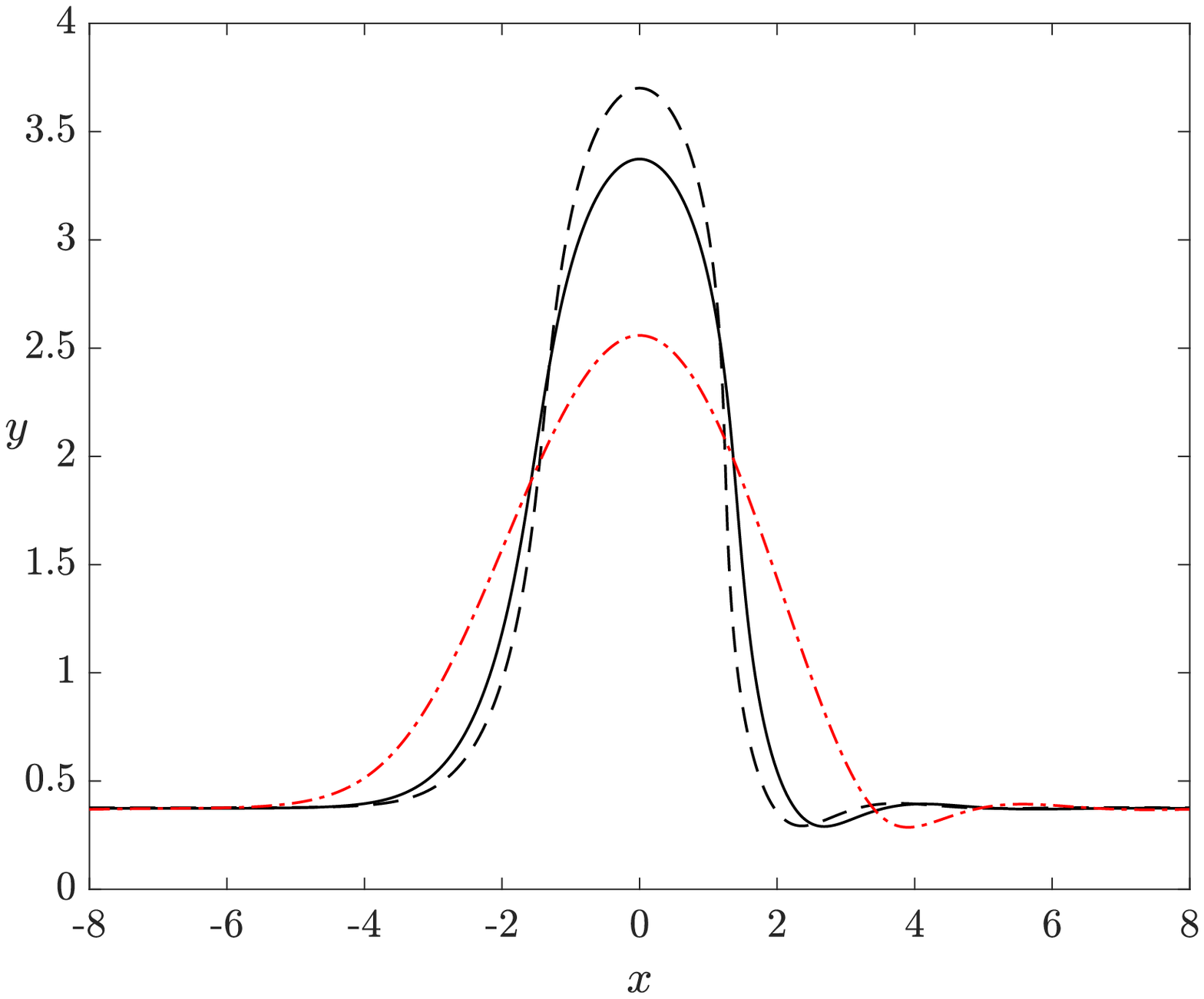}
\caption{}
\end{subfigure}
\end{center}
\caption{Fixed volume calculation with $\Ca=0.3$ and $V=15$ (so $\Ca V = 4.5$). Thin-film calculation for the FCM \eqref{eq:TFE2_full}, shown with solid lines, and the LCM \eqref{eq:TFE1}, shown with dot-dashed lines, and the Stokes calculation shown with a thick solid line, all on the domain $[-8,8]$. (a) Drop heights $H$ versus inclination angle $\beta$, including a close-up inset near to the turning point. (b) Maximum of the absolute value of the drop slope versus inclination angle $\beta$. (c) Drop profiles at $\beta=2.95$ corresponding to the filled and empty circles (solid and dashed lines, respectively, for the FCM) and the filled square (dot-dashed line for the LCM) in the diagrams (a) and (b).} 
\mylab{fig:YL-FCMcomp2}
\end{figure}
%

If the volume is increased so that $\Ca V = 4.5$, then there is no YL solution at $\beta=\pi$. In this case the bifurcation curves for the FCM and for Stokes flow both turn around before reaching $\beta=\pi$, as can be seen in figure~\ref{fig:YL-FCMcomp3}. The bifurcation curve for the LCM continues to $\beta=\pi$, as expected, and the solution profile approaches a pure cosine of support $(2/\Ca)^{1/2}\pi \approx 62.8$. The bifurcation curve for the FCM terminates when the slope at one point on the downstream side of the pulse becomes infinite (see panel b of the figure), thus indicating a breakdown of the model. Profiles for the FCM model on the lower and upper branches of the bifurcation curve can be seen in panel (c), including that near to the infinite slope singularity at $\beta=2.9439$; the corresponding LCM solution is shown at the same $\beta$ value. The breakdown of the FCM model appears to occur at the point where the profile is about to become multi-valued. This assertion is supported by the Stokes calculations.
In this case the bifurcation curve (thick blue line in panel b) passes through the point where the profile becomes multi-valued and we are ultimately forced to terminate the branch due to computational difficulties. The most extreme profile for Stokes flow, corresponding to the empty star symbol in panel (a), is shown as the solid curve in panel (d). It is striking that this wave profile closely resembles a hanging drop close to the point of pinch-off and subsequent dripping. 

For the larger Bond number case with $\Ca V = 2.4$ the bifurcation curves all tend towards $\beta=\pi$ as can be seen in panel (a) of figure~\ref{fig:YL-FCMcomp} where the profiles for the FCM and for Stokes flow tend to conform with the YL solution (panel b), and the limiting profile for the LCM is a cosine wave with support $(2/\Ca)^{1/2}\pi \approx 8.1$. Notably the FCM and the Stokes model agree well near to $\beta=\pi$, as might be expected, but show significant divergence for inclination angles away from horizontal. The wave speed, $c$, and the precursor thickness, $h_p$ both approach zero as $\beta\to \pi$; and for the FCM this occurs such that $c\sim 2.83(\pi-\beta)^{3/2}$ and $h_p\sim 1.80(\pi-\beta)$ as discussed in \S~\ref{appendix:asymp}. 
Figure~\ref{fig:chasym} shows a comparison between these asymptotic predictions for the FCM and the numerical results, with excellent agreement between the two. We have also confirmed that our numerics agree with the near-horizontal asymptotics for the LCM case which predicts according to \eqref{hpexp} that 
$h_p\sim  2.17(\pi-\beta)$.

For the same Bond number at larger volume so that $\Ca V = 4.5$, the results in figure~\ref{fig:YL-FCMcomp2} show that the bifurcation curves for the FCM and for Stokes flow have a turning point and the LCM curve approaches a pure cosine solution with support $8.1$ at $\beta=\pi$ (compare figure~\ref{fig:YL-FCMcomp3}). Beyond the turning point the bifurcation curve for the FCM terminates at an infinite-slope singularity (see panel (b) for detail), and the curve for Stokes flow stops because of computational issues.

%
\begin{figure}
%
%
\begin{subfigure}{0.5\textwidth}
\includegraphics[width=2.46in]{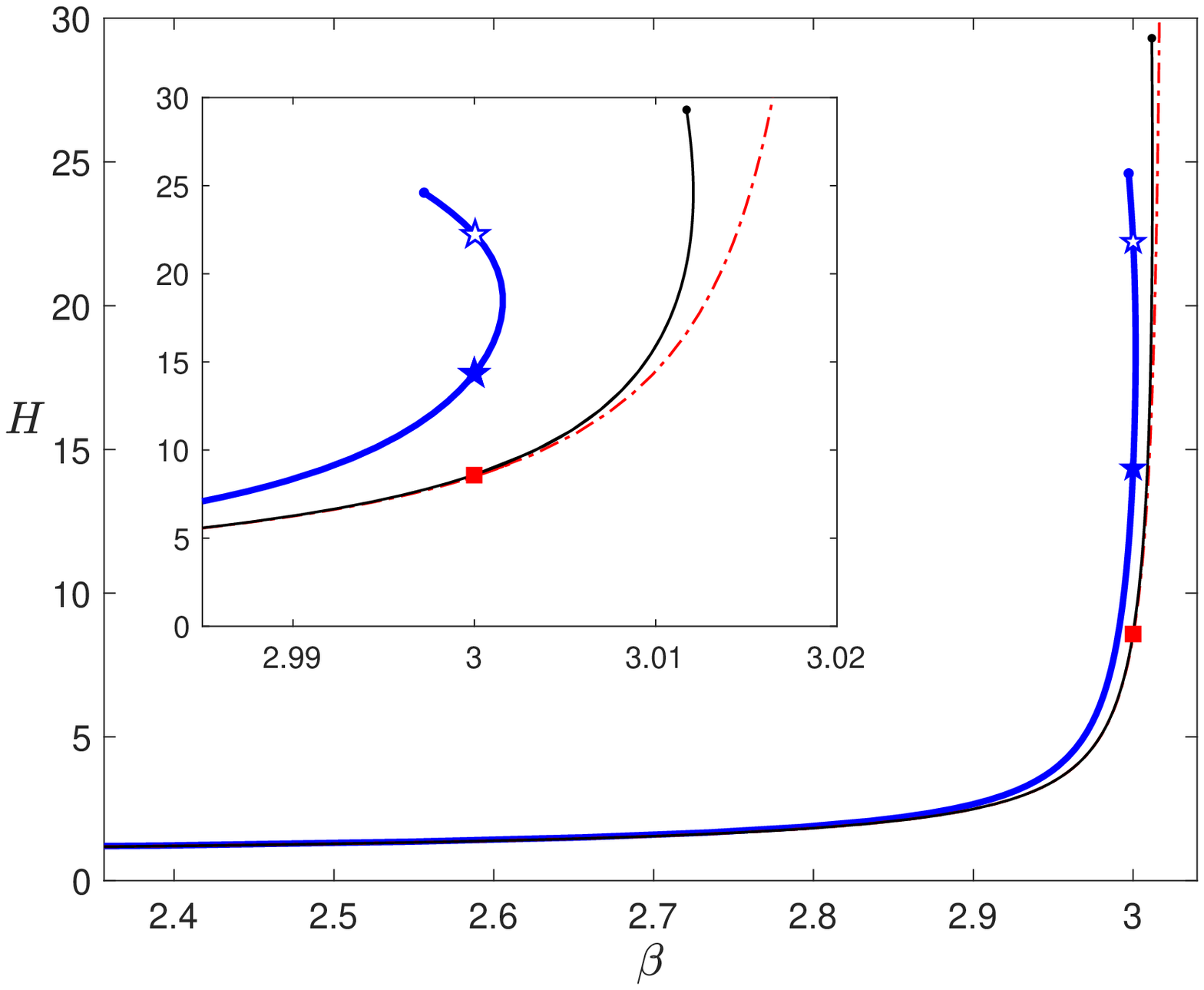}
\caption{}
\end{subfigure}
\begin{subfigure}{0.5\textwidth}
\includegraphics[width=2.46in]{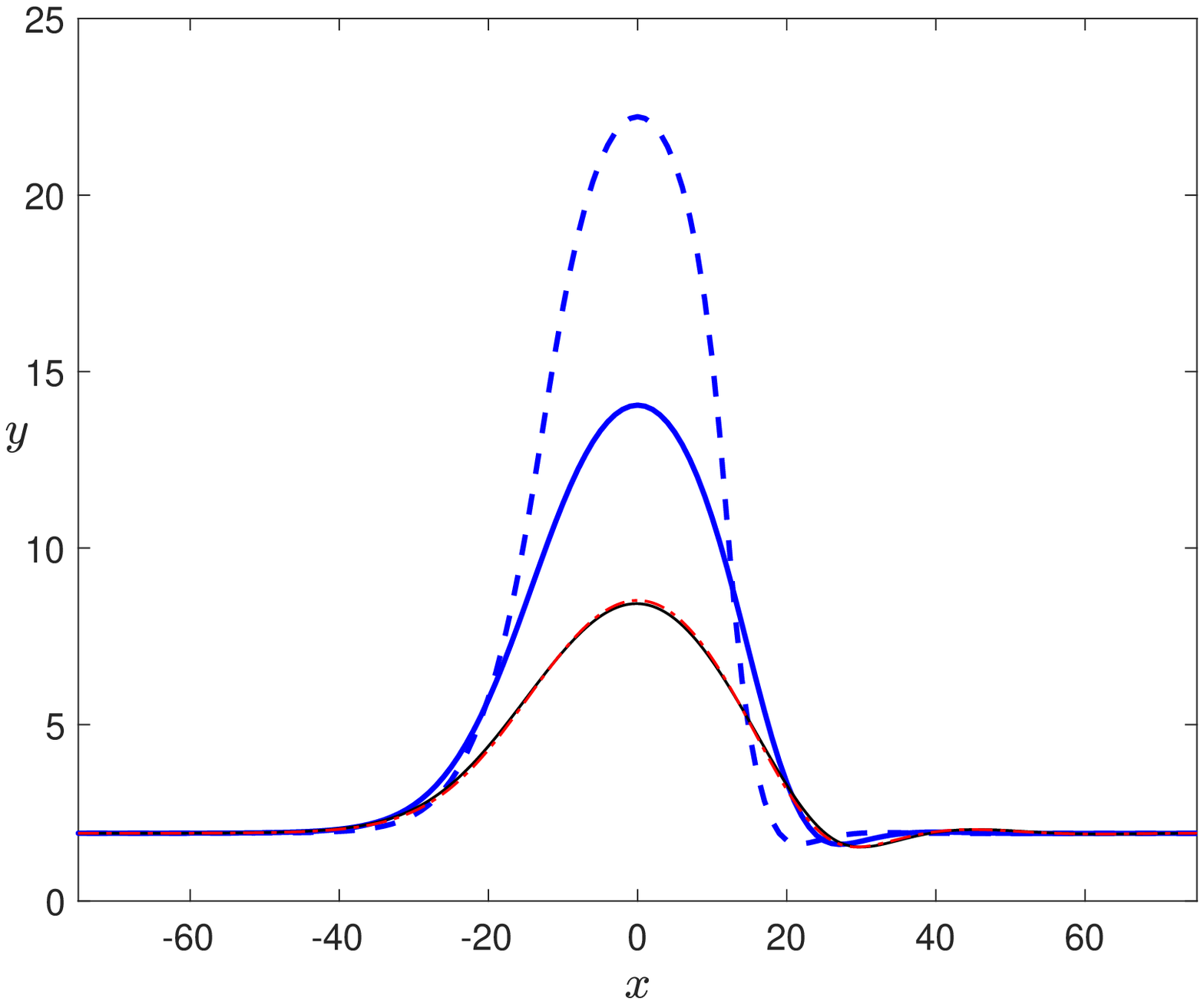}
\caption{}
\end{subfigure}
\caption{Fixed flow rate calculation with $\Ca=0.005$ and $q=2/3$. Thin film calculation for the FCM \eqref{eq:TFE2_full}, shown with thin solid lines, the LCM \eqref{eq:TFE1}, shown with dot-dashed lines, and the Stokes calculation, shown with thick solid lines. 
(a) Drop heights $H$ versus inclination angle $\beta$ with the inset showing a close-up. (b) Drop profiles at $\beta=3.0$. The profile on the lower branch for the boundary-integral Stokes calculation (filled star symbol in panel a) is shown with a thick solid line, and that on the upper branch (empty star symbol in panel a) is shown with a broken line. The almost coincident lowermost curves are the profiles for the LCM and FCM corresponding to the square symbol in panel a).
In (a) the LCM curve has a vertical asymptote at $\beta = 3.022$ according to \eqref{hpexp2}.}
\mylab{fig:thinStokes}
\end{figure}
\begin{figure}
\begin{subfigure}{0.5\textwidth}
\includegraphics[width=2.6in]{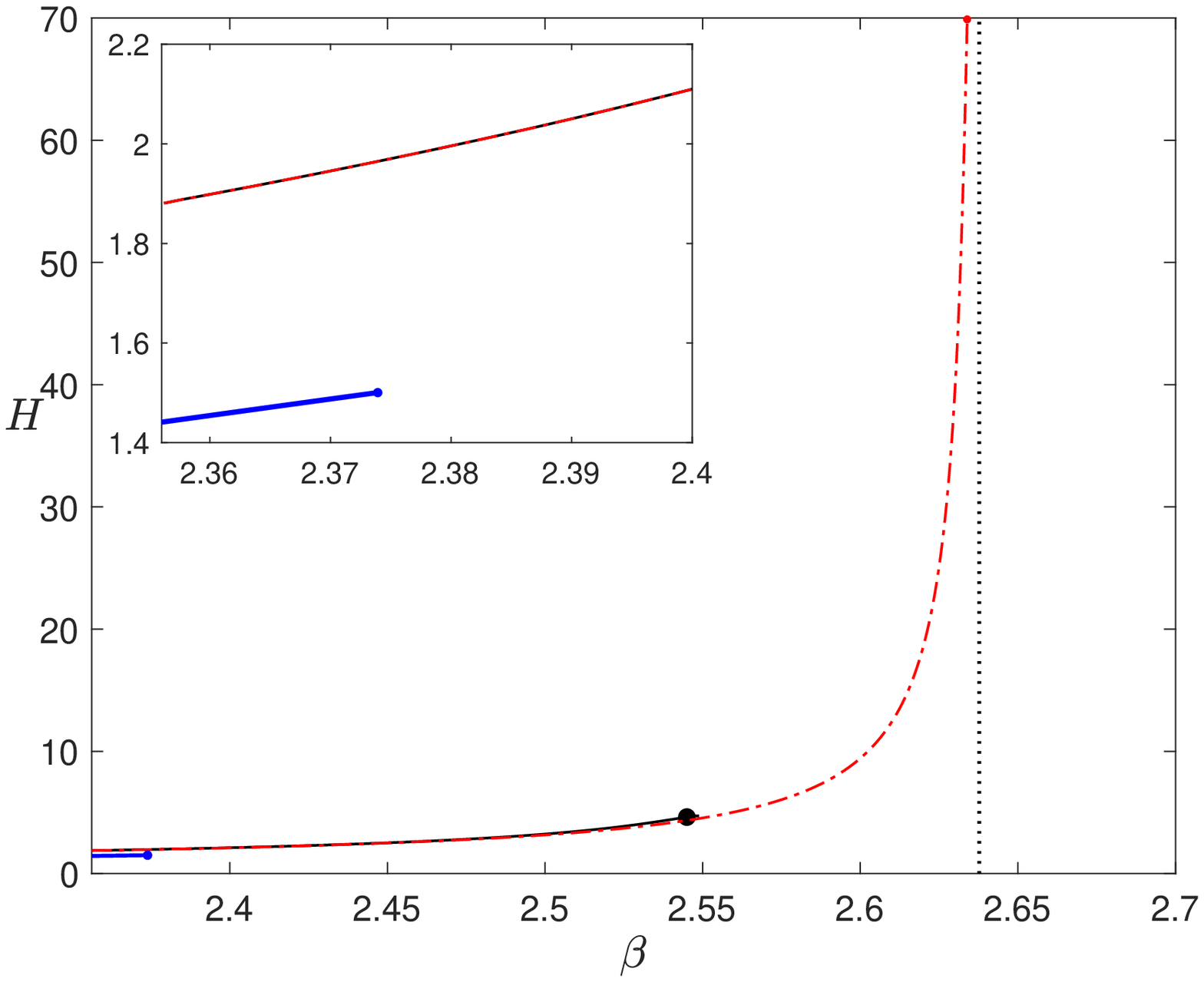}\hspace{0.025in}
\caption{}
\end{subfigure}
\begin{subfigure}{0.5\textwidth}\includegraphics[width=2.6in]{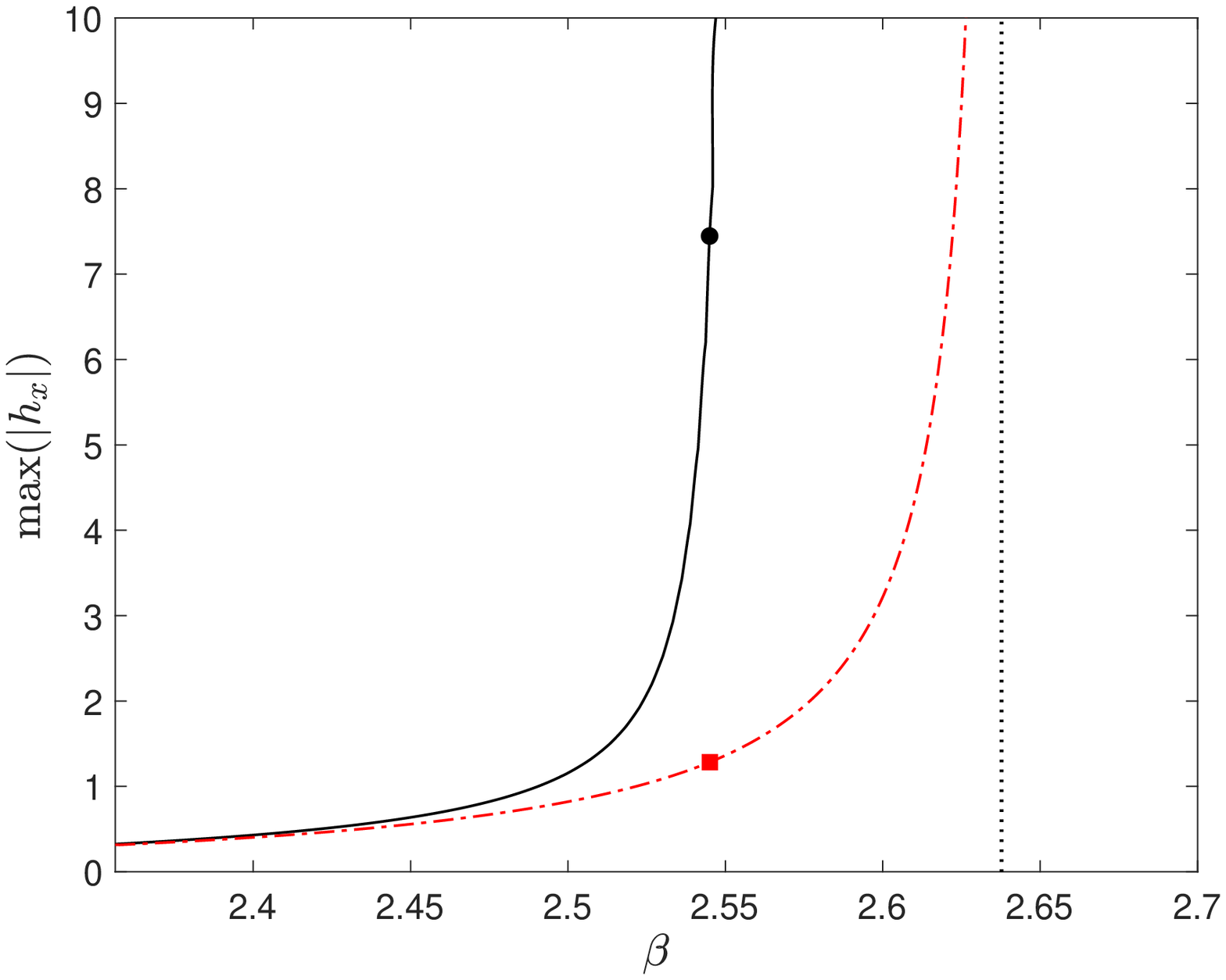}
\caption{}
\end{subfigure}
\begin{subfigure}{0.5\textwidth}
\includegraphics[width=2.6in]{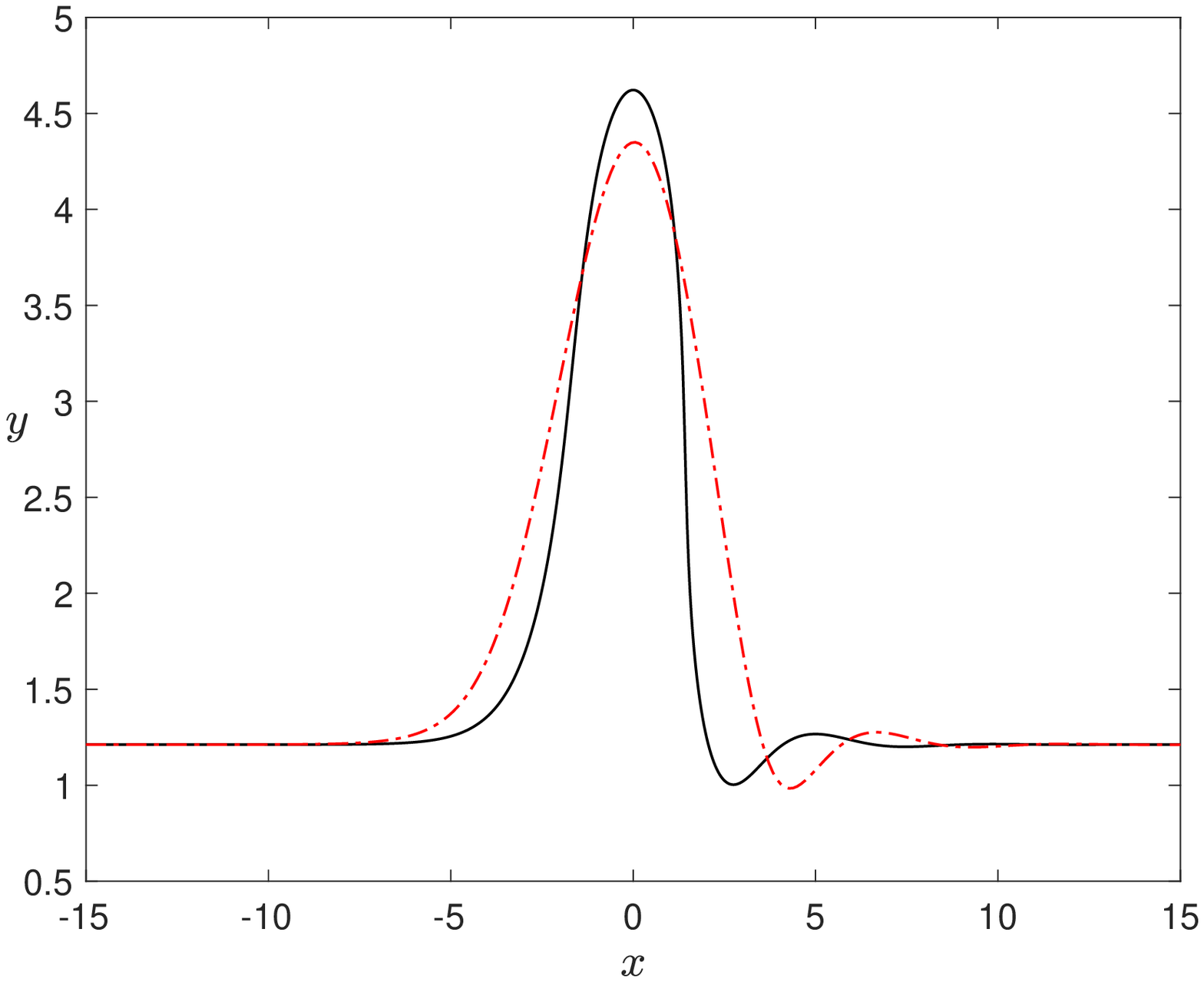}
\caption{}
\end{subfigure}
\begin{subfigure}{0.5\textwidth}
\includegraphics[width=2.6in]{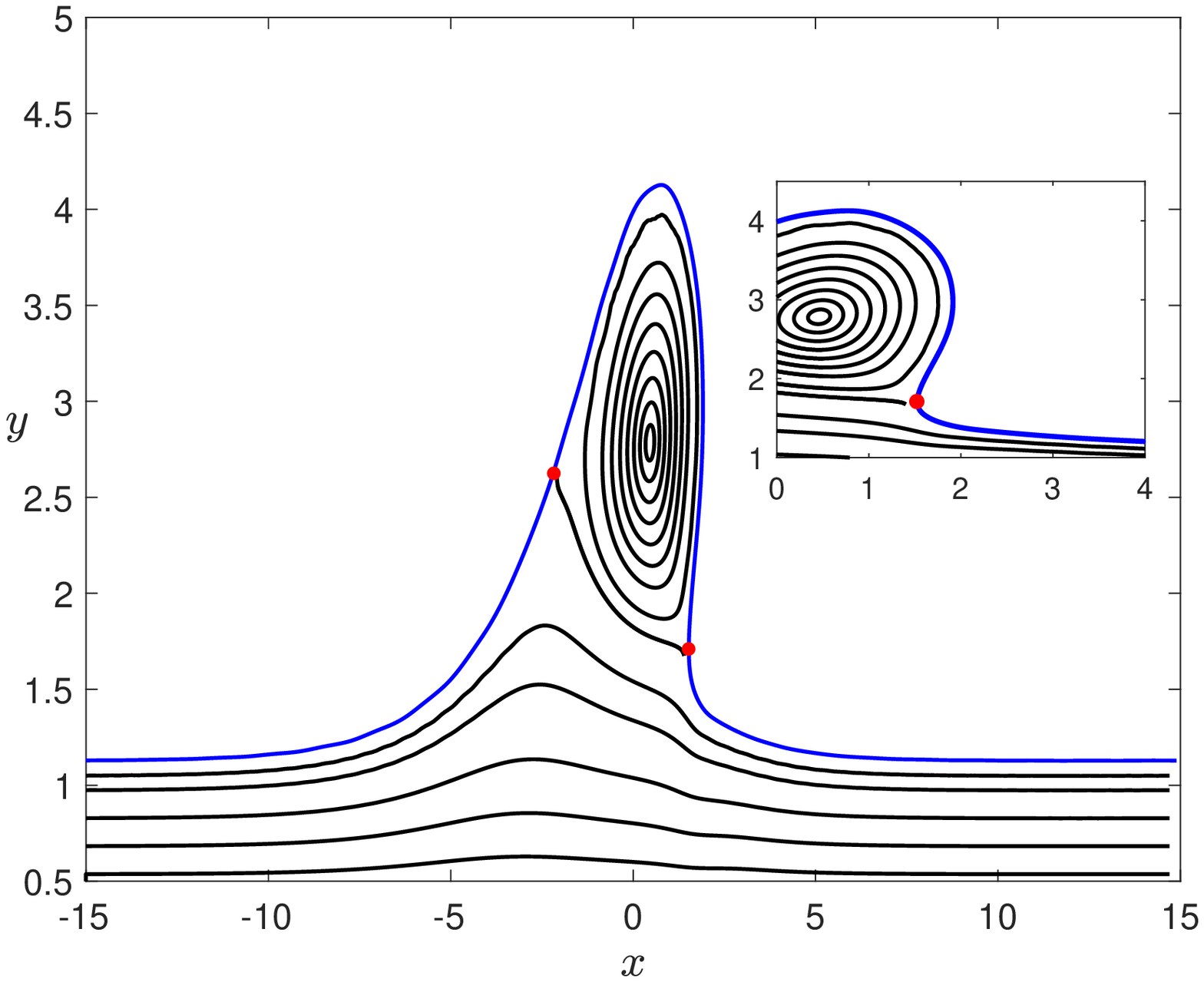}
\caption{}
\end{subfigure}
\caption{}
\caption{Fixed flow rate calculation with $\Ca=0.3$ and $q=2/3$.  The results for the FCM \eqref{eq:TFE2_full} and the LCM \eqref{eq:TFE1} are shown with thin solid lines/ dot-dashed lines respectively, and the Stokes calculation is shown with a thick solid line. 
(a) Drop heights $H$ versus inclination angle $\beta$. (b) Maximum of the absolute value of the drop slope versus inclination angle $\beta$. (c) Drop profiles at $\beta=2.545$ corresponding to the filled circle (solid line for the FCM) and the filled square (dot-dashed line for the LCM) in the diagrams (a) and (b). 
In (a) and (b) the vertical dotted line indicates the blow-up angle $\beta = 2.638$ (from \eqref{hpexp2}) for the LCM. (d) The overturning Stokes flow profile at $\beta=2.374$. The solid (red) dots indicate the stagnation points. The flow is from right to left below the eddy and in the clockwise direction inside the eddy.}
\mylab{fig:YL-FCM_fixed_q}
\end{figure}

\subsection{Fixed flow rate}

Turning now to the case of fixed flow rate, in figure~\ref{fig:thinStokes} we show results for the case of small Bond number, $\Ca=0.005$, for $q=2/3$. The drop height for the linear curvature model blows up at a critical inclination angle given by $\beta=3.022$ according to \eqref{hpexp2}. As was discussed in \S\ref{appendix:asymp} the near blow-up behaviour is described by the asymptotic analysis of \cite{kalliadasis1994drop} and \cite{yu2013velocity}. The LCM agrees well with the FCM for inclination angles up to about $\beta=3$. The LCM and FCM drop profiles are almost coincident at $\beta=3.0$, as can be seen in panel (b). The bifurcation curve for the FCM has a turning point at $\beta\approx 3.012$. The FCM predicts the correct qualitative behaviour but the location of the turning point is delayed compared to the Stokes calculation, where the turning point occurs at $\beta=3.002$. The bifurcation curve for the FCM computation terminates due to an infinite slope singularity (cf. figure~\ref{fig:YL-FCMcomp2}) and the bifurcation curve for Stokes flow stops where shown due to numerical difficulties. Overall, we see that \cbl for sufficiently small Bond number \cb the boundary of the travelling pulse solutions, that is the inclination angle beyond which such solutions do not exist, is consistently predicted to within only a small error by the two long-wave models. Therefore, if we infer that dripping occurs beyond this boundary, the LCM and the FCM both provide an accurate predictor for dripping. Furthermore, for the LCM we have a simple formula for the boundary value, given by \eqref{hpexp2}.

Results for fixed flow rate at the larger Bond number $\Ca=0.3$ and $q=2/3$ are shown in figure~\ref{fig:YL-FCM_fixed_q}. The bifurcation curve for the LCM pulse solutions exhibits blow-up at $\beta=2.638$, as predicted by \eqref{hpexp2}, and beyond this point such solutions do not exist. In this case it is more difficult to assess the consistency of this prediction with the FCM and the Stokes calculations. The FCM computation fails at an infinite-slope singularity that occurs earlier than the LCM blow-up point at $\beta\approx 2.55$. The Stokes profile shows strong overturning behaviour well before the LCM blow-up angle. Panel (d) shows the Stokes profile at $\beta=2.374$. Numerical convergence issues prevent us continuing the solution beyond this point.  For smaller inclination angles the eddy inside the pulse is relatively small and confined to the near-tip region. {\color{black} However, as the inclination angle increases and approaches the last computed point at $\beta=2.374$, the eddy grows in size and the rightmost stagnation point moves down toward the heel of the pulse (indicated by the solid dot in panel d of figure~\ref{fig:YL-FCM_fixed_q}). The curvature at the stagnation point grows substantially as the stagnation-point moves downwards leading us to conjecture that the free surface might be approaching a cusp at a critical $\beta$. Cusping in two-dimensional Stokes flows has been discussed by a number of authors \cite[e.g.][]{richardson1968two,jeong1992free}. 
Unfortunately, our present code is unable to reveal greater detail and further work is needed to determine what is going on in this region -- this is the subject of our ongoing investigations.}

\section{Summary and discussion}

We have examined the flow of a liquid film on the underside of an inclined flat plate in the absence of inertia with a view to describing the onset of dripping. In particular we have used several different 
model equations (a linearised and full curvature lubrication model, LCM and FCM, respectively, and the full equations of 
Stokes flow) to compute travelling-wave solutions on the assumption of either fixed volume in 
a periodic domain or else a constant flow rate. Our particular focus has been on following the 
solutions using parameter continuation toward the case where the plate is horizontal.

In this limit LCM and FCM solutions approach localised pure cosine and YL 
solutions, respectively, the latter only existing if the volume is smaller than a certain critical 
value. We have investigated the fixed-volume horizontal limit for the FCM model using an 
asymptotic analysis that generalises that presented by other authors for linearised curvature
\cite[see][]{yu2013velocity,kalliadasis1994drop}. For the FCM, if the volume is sufficiently large, 
the travelling-pulse solutions cease to exist beyond a certain critical plate inclination angle 
corresponding to a turning point in the bifurcation curve. Beyond the turning point the FCM 
eventually breaks down at an infinite slope singularity. For a single parameter set, and at fixed 
volume, \cite{Kofman_etal_2018} computed bifurcation curves for travelling-wave solutions 
using a number of different weighted residual integral boundary layer models that take inertia 
into account, and they also detected an infinite slope singularity, but it is unclear from their 
results whether a turning point is reached first. We suggest that it is the turning point which is 
connected with the onset of dripping.
For the fixed-flow-rate case, the LCM solutions grow in amplitude as the plate tends to become 
more horizontal, and they eventually blow up at a critical inclination angle, which is predicted by 
the asymptotic analysis. In contrast, the bifurcation diagrams for the FCM and Stokes flow 
models exhibit turning points at certain inclination angles. 

\begin{figure}
\centering
\includegraphics[scale=0.55]{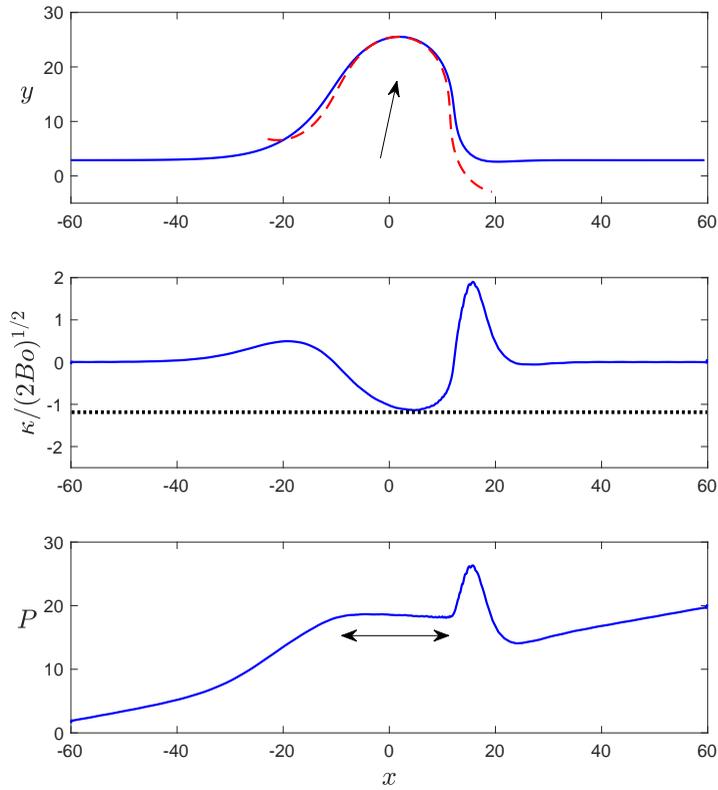}
%
%
%
\caption{Comparison with the static YL solution for the case of fixed-volume Stokes flow shown in figure~\ref{fig:YL-FCMcomp3}. The top panel shows the film profile (blue solid line) at the turning point, $\beta =2.9158$, with the YL solution computed at $\beta=\pi$ and then rotated through an angle $\pi-2.9158$. The arrow indicates the direction of gravity. The middle panel shows the scaled film curvature, $\kappa/(2\Ca)^{1/2}$ (the dotted line indicates the tip curvature for the maximum volume YL solution). The bottom panel shows $P$ defined in \eqref{eq:TFE2_full} which represents the combined effect of hydrostatic and capillary forces. The arrow in the bottom panel indicates the region over which the YL equation approximately holds.}
\mylab{fig:YLcomparison}
\end{figure}

\cite{zhou2022dripping} discussed dripping of a liquid film from the underside of a flat plate. 
They performed numerical computations in the presence of fluid inertia using the open-source finite-volume software Basilisk (http:// basilisk.fr). By prescribing the fluid volume over a periodic domain and then slowly increasing the inclination angle during an unsteady simulation, they were able to obtain quasi-equilibria corresponding to travelling wave solutions of the type we have computed here. They placed special emphasis on the film curvature at the tip of the wave crest, $\kappa_T$ (here tip refers to the local maximum at the wave crest with respect to a set of coordinates where the $x$-axis is horizontal and the $y$-axis points vertically downward in the direction of gravity). They suggested that the onset of dripping essentially starts at that point in time at which the tip curvature exceeds (in absolute value) the value $|\kappa_c|$ which obtains for the maximum-volume YL profile for a film underneath a horizontal plate (i.e. that found at the turning point in our figure~\ref{fig:YLsoln}). To reconcile this observation with our results, we 
show in figure~\ref{fig:YLcomparison} a comparison between one of our fixed-volume travelling-wave solutions and solutions of the YL equation. The upper panel shows the travelling wave calculated at the turning point of the bifurcation curve for Stokes flow in figure~\ref{fig:YL-FCMcomp3} (shown there with a thick solid line; the turning point occurs at $\beta=2.9158$). Since, as was noted by \cite{zhou2022dripping}, the axis of the drop shape over the main part of the pulse is almost aligned with the direction of gravity, we superimpose onto the profile the YL solution computed at $\beta=\pi$ for the same Bond number (the YL profile has been rotated through an angle $\pi-\beta$ so that the gravity vectors, indicated by the arrow, for the two solutions are aligned.) The middle panel shows the curvature of the wave profile together with the value for appropriately scaled tip curvature of the maximum-volume YL solution. Evidently the tip curvature of the travelling pulse is very close to that for the YL solution corroborating the observation made by \cite{zhou2022dripping}. However, if we examine the force balance at the surface, shown in the bottom panel, we see that the YL equation is only relevant over a portion of the domain (that, incidentally, includes the wave tip), this portion being that where $P$ is approximately constant, indicated by the arrow. Here $P$ as defined in \eqref{eq:TFE2_full} corresponds to the force balance in the YL equation for a drop hanging underneath a horizontal wall. The fact that $P$ is not constant over other parts of the wave indicates that the flow within the pulse has an important effect in shaping the drop profile. Nevertheless the observation that the tip curvature reaches the same value as the maximum volume YL tip curvature at the turning point of our bifurcation curves is interesting. 

Using the tip curvature as an indicator of dripping is perhaps not so useful in practice. However, since we have now established that the maximum-volume YL tip curvature occurs at the turning point of our bifurcation diagrams, we may instead use these as a practically useful indicator of dripping. They supply an estimate for the inclination angle at which dripping will start to occur.

For small Bond number the inclination angle that we associate with dripping onset is consistently predicted to within only a small error by the thin-film models. Therefore, if we accept that dripping occurs for more extreme inclination angles, the LCM and the FCM both provide a useful prediction for dripping transition. Given their relative simplicity and amenability to straightforward numerical computation, they therefore offer a relatively simple and potentially effective tool for dripping prediction. At larger Bond number, the lubrication models (LCM and FCM) are qualitatively but not quantitatively accurate and full Stokes calculations are needed.

Finally, we note that for the fixed flow rate case, when the Bond number is relatively large, the solutions of the LCM blow up as before, but the computations for both the FCM and the Stokes models break down before a turning point is reached. The FCM fails at an infinite slope singularity indicating that the profiles are tending to become multi-valued. However, the reason for the breakdown of the Stokes model is unclear, and this remains as a topic for future investigation.

\vspace{0.125in}
\noindent
{\bf Declaration of Interests}. The authors report no conflict of interest.


\appendix

\section{Static drops}\label{sec:static}

When the inclination angle $\beta=\pi$, the liquid film is static and its surface shape is described by the static 
YL equation. 
According to the scales introduced in \S\ref{sec:problem}
this takes the dimensionless form
\bea
\label{eq:YL0} 
\frac{\kappa}{\Ca} + 2y = P_0,
\eea
where $P_0$ is an {\it a priori} unknown constant reference pressure inside the drop. 
The first term in \eqref{eq:YL0} represents the capillary pressure due to surface curvature and the second term represents the hydrostatic pressure at the surface. Assuming that the drop meets the wall at a zero contact angle, a global force balance over the drop in the vertical direction shows that $P_0 = V/\ell$, where $2\ell$ is the support of the drop. Written in terms of the surface parameterisation $(a(s),b(s))$ shown in figure~\ref{YLfig}, the YL equation is then given by
%
\begin{figure}
\centering
\includegraphics[width=2.6in]{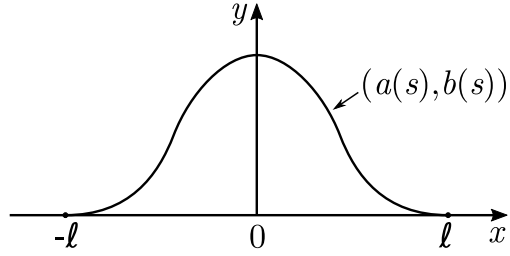}
\caption{(Color online) Sketch of the YL problem for a static drop under a horizontal wall ($\beta=\pi$) with support $2\ell$. In the sketch gravity acts in the positive $y$ direction.
}
\mylab{YLfig}
\end{figure}
%
\bea
\frac{1}{\Ca} \frac{a'b'' - b'a''}{(a'^2+b'^2)^{3/2}} + 2 b = \frac{V}{\ell}, 
\label{eq:YLnondim} 
\eea
In the case when the parameter $s$ represents arc length the denominator in the first term in \eqref{eq:YLnondim} is equal to unity. The drop is symmetric about $x=0$ and has support $2\ell$. 

\cite{pitts1973stability} obtained an exact solution to \eqref{eq:YLnondim} for general contact angle and for a given drop volume using elliptic functions. Of interest here is the case of zero contact angle since such solutions represent the limiting weak solutions that are approached in our travelling-wave calculations as $\beta\to \pi-$ and the wall tends to become horizontal. We have obtained solutions of this type numerically using the procedure described in Appendix~\ref{appendix:static} and confirmed that they agree with those of \cite{pitts1973stability}.
%
\begin{figure}
\begin{subfigure}{0.5\textwidth}
\includegraphics[width=2.6in]{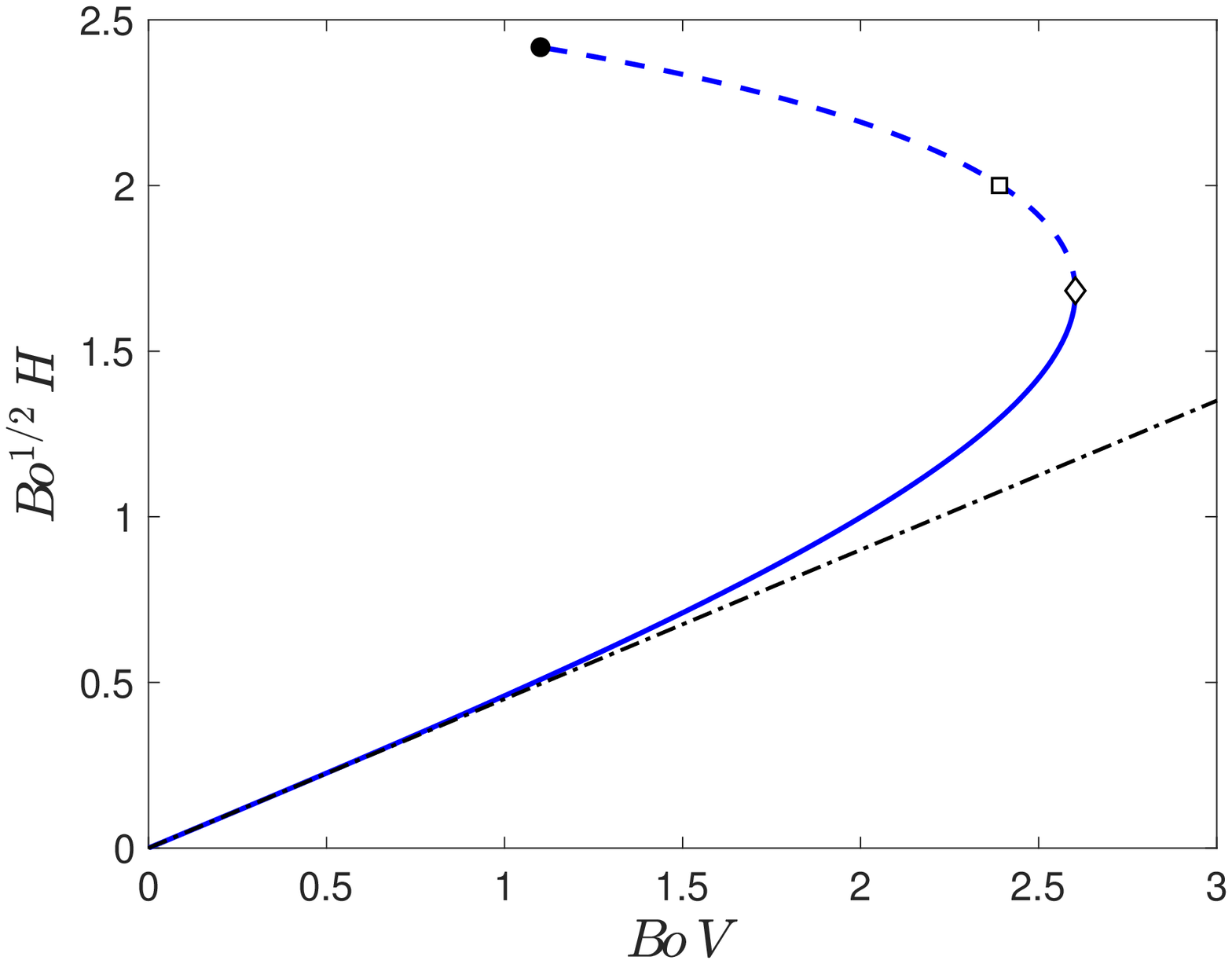}
\caption{}
\end{subfigure}
\begin{subfigure}{0.5\textwidth}
\includegraphics[width=2.6in]{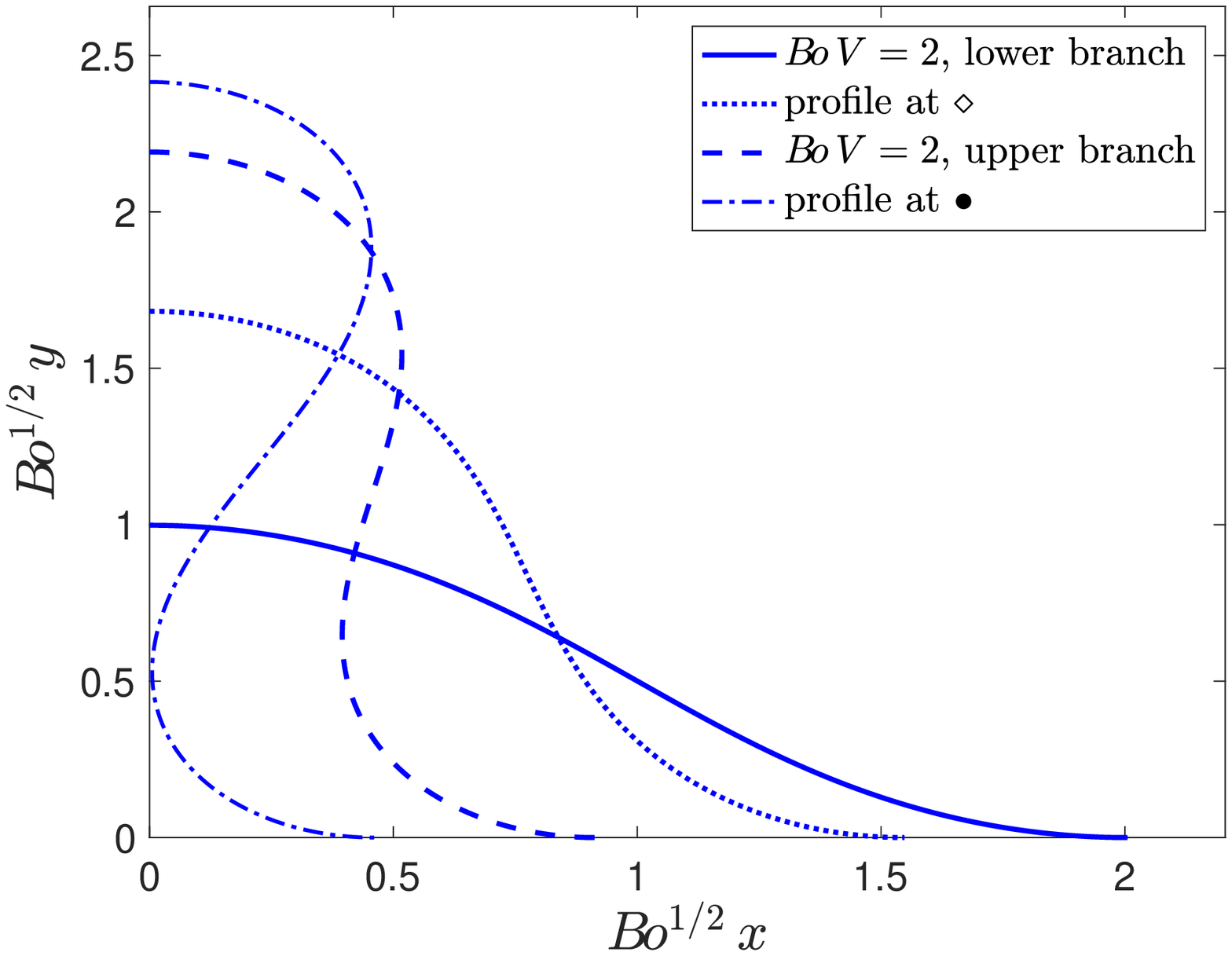}
\caption{}
\end{subfigure}
\caption{(Color online) Static drop solutions to the YL problem \eqref{eq:YLnondim}, \eqref{static_bcs} with volume constraint \eqref{static:volconstr}. (a) Bifurcation diagram showing scaled drop height $\Ca^{1/2} H$ against scaled drop volume $\Ca\,V$. The square symbol ({\tiny $\square$}) indicates the onset of multivaluedness in the solution profiles on the upper branch for $V<2.39\,\Ca^{-1}$. Solutions on the solid part of the curve are stable and on the broken part are unstable according to \cite{pitts1973stability}. The dot-dashed line corresponds to the linearised curvature solution to \eqref{YLlin}. (b) Drop profiles for particular volumes $V$ including the profile at the saddle-node ($\diamond$) where $V=2.60\,\Ca^{-1}$, and at the point of pinching ($\bullet$) where $V=1.10\,\Ca^{-1}$.}
\mylab{fig:YLsoln}
\end{figure}
%
\begin{figure}
\begin{subfigure}{0.5\textwidth}
\includegraphics[width=2.6in]{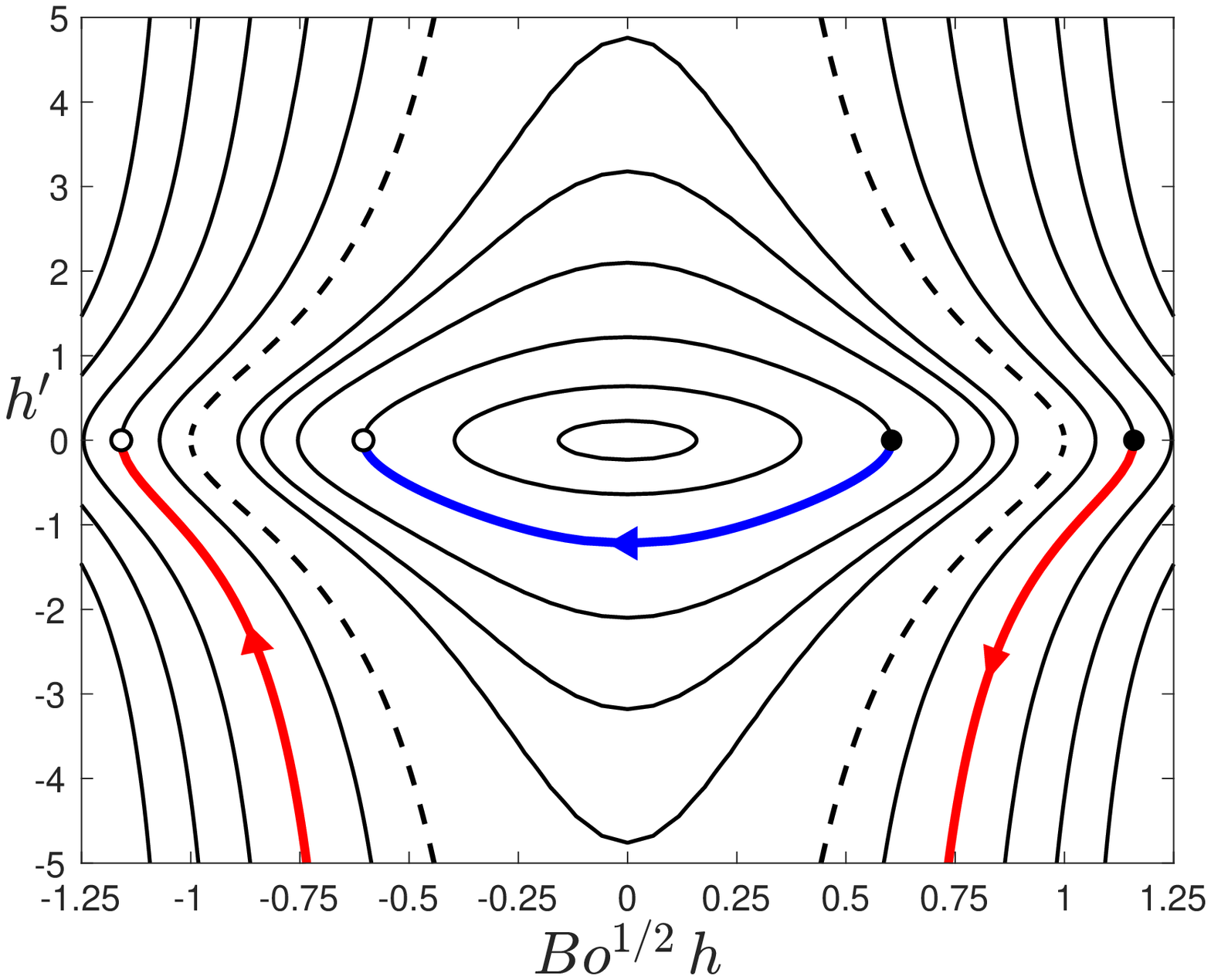}
\caption{}
\end{subfigure}
\begin{subfigure}{0.5\textwidth}
\includegraphics[width=2.6in]{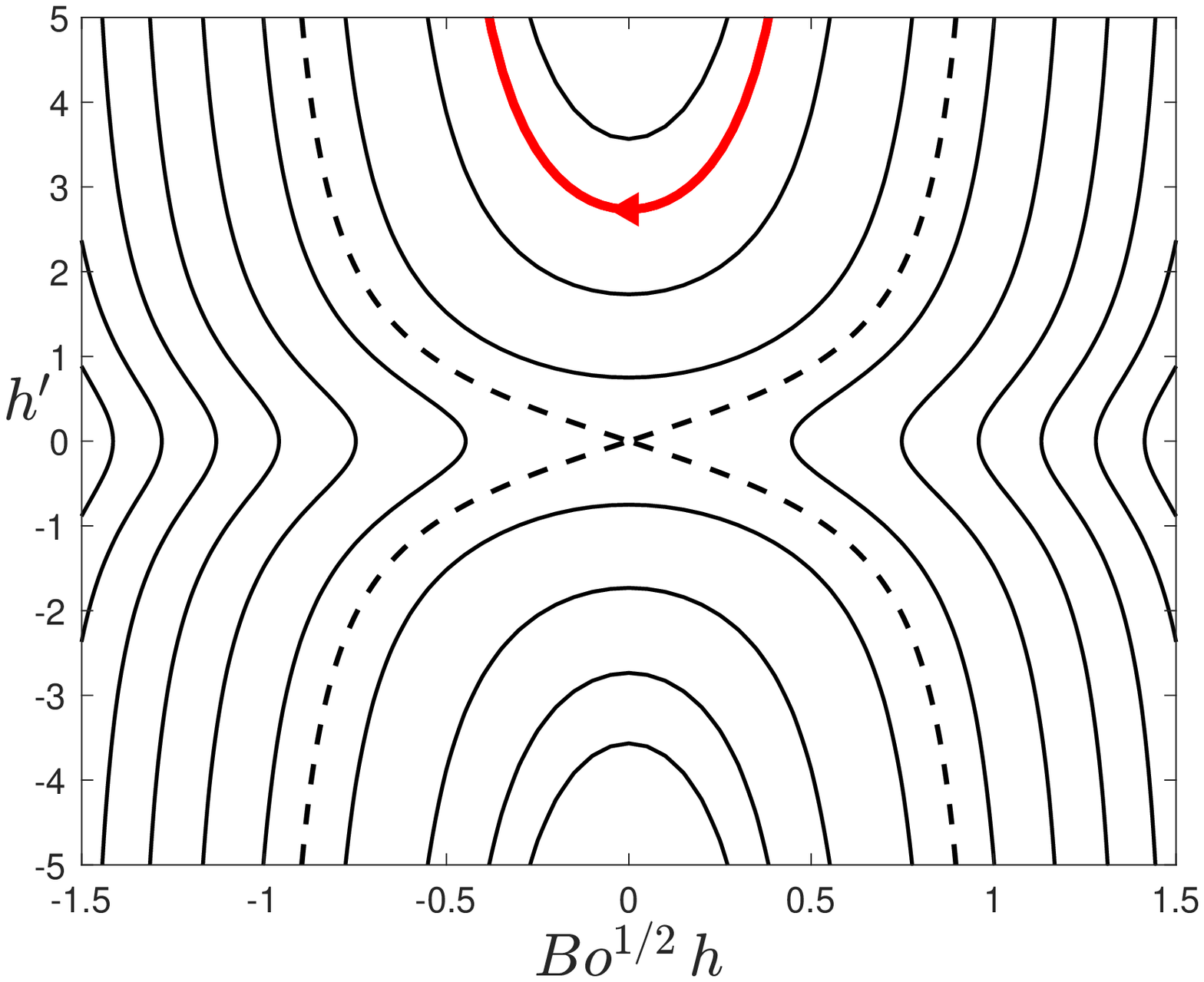}
\caption{}
\end{subfigure}
\caption{(Color online) Phase portraits in the $(h,h')$ plane for \eqref{singlesign_heq} with (a) the minus sign and (b) the plus sign, both shown for $\nu=0$. The trajectories correspond to different values of $E$: in (a) the dot at the origin corresponds to $E=-1$, the trajectory $E=0$ is dashed, and the closed orbits correspond to $-1\leq E\leq 0$; in (b) the trajectory $E=1$ is dashed, the $U$-shaped trajectories are for $0\leq E\leq 1$ and the remainder are for $E>1$). The thick blue and thick red trajectories correspond to one half of a single-valued and a multi-valued free-surface profile respectively. The filled circle indicates the drop maximum and the empty circle indicates the point of contact with the wall. In each case the phase portrait should be shifted by an appropriate choice of $\nu$ to move the empty circle to the origin. For the multi-valued profile a jump is made to the trajectory with the same value of $E$ in portrait in panel (b) and back again.}
\mylab{fig:pplane}
\end{figure}
%
The results are shown in figure~\ref{fig:YLsoln}. Since the Bond number can be removed from the static problem by scaling lengths by $\Ca^{-1/2}$, the bifurcation diagram in panel (a) shows the scaled dimensionless drop height $\Ca^{1/2}H$ against the scaled dimensionless drop volume $\Ca V$ (see the blue solid and blue dashed lines) . In addition, the black dot-dashed line corresponds to linearised curvature solutions, which will be discussed later.
The right panel 
shows drop profiles at particular values of $V$ along the bifurcation curve in the left panel.
The profiles are multivalued on the upper branch 
when $V<2.39\,\Ca^{-1}$. The presence of a saddle-node at $V=2.60\,\Ca^{-1}$ suggests 
that the solutions on the upper branch are unstable and those on the lower branch are 
stable. This is confirmed for `static' stability by the work of \cite{pitts1973stability} (we draw a distinction here between static and dynamic stability in the sense defined by \cite{lowry1995capillary}; the issue of dynamic stability for pendent drops was examined by \cite{pozrikidis2012stability} under conditions of Stokes flow, but it is not addressed here for our specific configuration).
The drop profiles on the upper branch pinch at $V=1.10\,\Ca^{-1}$. The upper 
branch can be continued beyond this point up to $V=0$ but all of the profiles are 
self-intersecting and therefore they are not physically relevant.

The existence of a turning point in the bifurcation curve indicates that there is no static 
solution for $V>2.60\,\Ca^{-1}$. Physical intuition suggests that beyond this point the drop volume is too large to be sustained by surface tension and dripping will ensue.

The preceding results can be explained using a phase plane analysis. Setting $s=x$, $a=\pm x$ and $b=h(x)$ in \eqref{eq:YLnondim}, we obtain a form appropriate for computing a single-valued solution $h(x)$. Integrating this equation once we obtain
\bea \label{singlesign_heq}
\mp\frac{1}{(1+h'^2)^{1/2}} + (\Ca^{1/2}\,h - \nu)^2 = E,
\eea
where $\nu = \Ca^{1/2}P_0/2$ and $E$ is a constant of integration.
The $-/+$ sign corresponds to the case when $\zh{n}\cdot \zh{j}$ is negative/positive, where 
$\zh{j}$ is the unit vector in the $y$ direction. The phase portraits for either choice of sign are shown in figure~\ref{fig:pplane} for the case $\nu=0$: the constant $\nu$ can be effectively removed from \eqref{singlesign_heq} via the mapping $h\mapsto h + \Ca^{-1/2}\nu$, which corresponds to a horizontal translation of the phase portrait. A single-valued or a multivalued drop profile is constructed by following the thick blue or red trajectories. In both cases the profiles are symmetric with respect to the inflection point at $x=\ell/2$. For a multivalued drop, a jump is made from the portrait in panel (a) to the trajectory with the same value of $E$ in panel (b), and then back again.
In both cases an appropriate horizontal shift is required to ensure $h=0$ at the point of contact with the wall.  

In the case of the linearised curvature approximation, the reduced form of the YL equation is
\bea \label{YLlin}
\frac{1}{\Ca} h_{xx} + 2h = \frac{V}{\ell}.
\eea
It has the solution $h = (V/2\ell)(\cos \sqrt{2\Ca}\,x+1)$ for $|x|<\ell$ where $\ell =  \pi/\sqrt{2\Ca}$. This is represented by the straight dot-dashed line with slope $\sqrt{2}/\pi$ in the left panel of figure~\ref{fig:YLsoln}.

\section{Numerical solution of the YL equation}\label{appendix:static}

Assuming left-right symmetry, we solve \eqref{eq:YLnondim} subject to the boundary conditions
\begin{subequations} \label{static_bcs}
\begin{align}
&& & a(0) = 0, & a'(0)=1,&      &b(0)=H,&    & b'(0) = 0, & && \\
&& & a(S) = \ell, & a'(S)=1,&  & b(S)=0,\, \, &    & b'(S) = 0,& &&
\end{align}
\end{subequations}
where with reference to the sketch in figure \ref{YLfig} the support of the drop is of length $2\ell$, $H$ is the height at the 
centre of the drop, and $2S$ is the total arclength over the drop surface. 
The derivative boundary conditions in \eqref{static_bcs} assume zero contact angle at the wall, as is illustrated in figure \ref{YLfig}. Such solutions represent the limiting weak solutions that are approached in our travelling-wave calculations as $\beta\to \pi$ and the wall tends to become horizontal. 
The lengths $\ell$ and $H$ are to be determined as part of the solution to the problem subject to the 
constraint
\bea \label{static:volconstr}
\int_0^S ba' \dd s = \frac{V}{2},
\eea
which fixes the volume of the drop to be $V$. 

Here we obtain a solution numerically by first rewriting \eqref{eq:YLnondim} as the first-order system. $\bm{u}' = \bm{F}(\bm{u})$, where $\bm{u} = (u_1,u_2,u_3,u_4) = (a,a',b,b')$, and
\bea
\bm{F} = \left (u_2,\, \Ca (2u_3u_4 - u_4 V/\ell),\, u_4,\, \Ca (-2u_2u_3 +u_2 V/\ell) \right ).
\eea
The second and fourth entries in $\bm{F}$ have been obtained by multiplying \eqref{eq:YLnondim} by $b'$ and $a'$ respectively, and then using the fact that $a'^2+b'^2=1$. 
We can eliminate the Bond number from \eqref{eq:YLnondim} by making the transformation $s\mapsto \Ca^{-1/2}s$, $a\mapsto \Ca^{-1/2} a$, $b \mapsto \Ca^{-1/2} b$ and $A \mapsto \Ca^{-1/2}A$. It is therefore sufficient to solve for the case $\Ca=1$;  solutions for other Bond numbers can be obtained by using the aforementioned rescaling. In numerical practice, we guess the values of $H$ and $\ell$ and then shoot forwards from $s=0$, using Runge-Kutta integration for example, until $b'=0$. We then compute the volume so obtained and refine the values of $H$ and $\ell$ using Newton iterations until the drop volume attains the desired value $V$, and $b$ vanishes at the same point as $b'$. The value of $S$ is extracted from the converged solution. Having obtained a solution for one value of $V$, we use arclength continuation to follow the solution branch as $V$ changes.


\section{Linear stability of a flat film for Stokes flow}
\label{appendixstokesflat}

In this appendix we present a brief discussion of the linear stability of a flat film under conditions of Stokes flow. We work under the same non-dimensionalisation introduced in \S\ref{sec:problem}.

We perturb the Nusselt solution corresponding to a flat film of unit dimensionless thickness by
introducing a small disturbance, writing
\bea \label{cats}
h(x,t) = 1 + A \left(\ee^{\sigma t}\,\ee^{\ri kx} + \mbox{c.c.}\right),
\eea
where $A \ll 1$, $k$ is the real wave number of the disturbance, $\sigma$ is the complex growth rate, and c.c. denotes the complex conjugate.
Working with a stream function, $\psi$, we make the expansion
\bea \label{expans}
\psi = (y^2-y^3/3)\sin \beta + A \psi_1(y)\ee^{\sigma t}\ee^{\ri k x} + \cdots.
\eea
Substituting into the Stokes governing equation $\nabla^4 \psi = 0$, and linearising, we obtain
\bea \label{orrsom}
\psi_1^{(iv)} - 2k^2 \psi_1'' + k^4 \psi_1 = 0,
\eea
where a prime denotes differentiation with respect to $y$.
To satisfy the no-slip and impermeability conditions,
we require that $\psi_1=\psi_1'=0$ at $y=0$.
Linearising the dimensionless form of the surface conditions (\ref{grapes}), we derive  the linearised tangential stress condition,
\bea \label{tang}
\psi_1''(1) + k^2\psi_1(1) = 2\sin \beta,
\eea
and the linearised normal stress condition
\bea \label{norm}
-\ri \psi_1'''(1) - 3k^2 \psi_1'(1) =  2k (\cos \beta-kWe)+ k^3/\Ca .
\eea
The linearised kinematic equation \eqref{banana2} yields
\bea \label{kin}
\psi_1(1) = \ri \sigma/k -\sin \beta.
\eea
The general solution to (\ref{orrsom}) is
\bea
\psi_1(\yh) = a_1\cosh(k y) + a_2y \cosh(ky) + a_3\sinh(ky) + a_4y\sinh(ky).
\eea
Compiling the boundary conditions, we assemble the
matrix system $\boldsymbol{A}\, \boldsymbol{x}=\boldsymbol{b}$, where
\bea
\hspace{-0.2in}
\boldsymbol{A}=
\left(
\begin{array}{cccc}
0 & 1 & k & 0 \\
1 & 0 & 0 & 0 \\
2k^2 \cosh k & 2k(\sinh k + k\cosh k) & 2k^2 \sinh k &  2k(\cosh k + k\sinh k) \\
2\ri k^3 \sinh k & 2\ri k^3 \sinh k & 2\ri k^3 \cosh k & 2\ri k^3 \cosh k
\end{array}
\right ), \eea and $\boldsymbol{x}=(a_1,a_2,a_3,a_4)^T$ and
$\boldsymbol{b}=(0,0,2\sin \beta,kT)$, where $T(k) = 2\cos \beta +
k^2/\Ca$. Substituting the solution to this system into \eqref{kin} we obtain
the growth rate \bea \label{app_eq_sig} \sigma = \left( \frac{k-\sinh k \cosh
k}{k^2+\cosh^2k}\right)\!\frac{T(k)}{2k} - \ri k \left(\frac{1 + \cosh^2 k +
k^2}{k^2+\cosh^2k}\right)\sin \beta. \eea Since $k-\sinh k \cosh k< 0$ for
all $k>0$, it follows that $\mbox{Re}(\sigma) >0$ if $T(k)<0$. This occurs
when $\cos \beta<0$ in which case $\mathrm{Re}(\sigma)>0$ if $k\in
(0,k_c)$ where
\bea \label{app:stokes}
k_c = \sqrt{-2\Ca \cos \beta}.
\eea
Accordingly, the cut-off wave number for linear instability under
conditions of Stokes flow is identical to that obtained for the LCM and FCM \cite[see][]{blyth2018two}.

\bibliographystyle{jfm}
\bibliography{xbib}

\end{document}